# Observation of generalized *t-J* spin dynamics with tunable dipolar interactions

Annette N. Carroll[1*], Henrik Hirzler[1], Calder Miller[1], David Wellnitz[1], Sean R. Muleady[1,2], Junyu Lin[1], Krzysztof P. Zamarski[1,3], Reuben R. W. Wang[1], John L. Bohn[1], Ana Maria Rey[1], Jun Ye[1*]

[1] JILA, National Institute of Standards and Technology, and Department of Physics, University of Colorado, Boulder, CO 80309, USA

[2] Joint Center for Quantum Information and Computer Science, Joint Quantum Institute, National Institute of Standards and Technology, and University of Maryland, College Park, MD 20742, USA

[3] Universität Innsbruck, Institut für Experimentalphysik und Zentrum für Quantenphysik, 6020 Innsbruck, Austria

[*]Corresponding authors. Emails: annette.carroll@colorado.edu (A.N.C) and ye@jila.colorado.edu (J.Y.)

**Abstract:** Long-range and anisotropic dipolar interactions profoundly modify the dynamics of particles hopping in a periodic lattice potential. We report the realization of a generalized *t-J* model with dipolar interactions using a system of ultracold fermionic molecules with spin encoded in the two lowest rotational states. We independently tuned the dipolar Ising and spin-exchange couplings and the molecular motion and studied their interplay on coherent spin dynamics. Using Ramsey spectroscopy, we observed and modeled interaction-driven contrast decay that depends strongly both on the strength of the anisotropy between Ising and spin-exchange couplings and on motion. This study paves the way for future exploration of kinetic spin dynamics and quantum magnetism with highly tunable molecular platforms in regimes that are challenging for existing numerical and analytical methods.

**Main Text:**

The *t-J* model, which arises from the more general Fermi-Hubbard model in the limit of large onsite interactions $U$, fundamentally describes a competition between motion via tunneling between lattice sites, $t$, and spin interactions, $J$, caused by superexchange via virtual tunneling of particles in adjacent lattice sites. The model has long been proposed to explain unconventional phases, including high-temperature superconductivity (*1–3*). Ultracold atoms with contact interactions have recently emerged as a controllable system to study the behavior of this model, but the $J{\sim}t^2/U$ scaling of superexchange has limited the range of accessible phases to $J \ll t$ and imposes stringent low temperature requirements to observe interaction-induced phase transitions (*4–6*).

Adding long-range anisotropic dipolar interactions, where $J$ can be tuned independently of $t$, softens the low temperature constraints on phase transitions and generates a generalized *t-J* model that is predicted to produce exotic phases including an enhanced superfluid state (*7*, *8*). Magnetic atoms provide access to dipolar interactions which, while relatively weak, have recently produced a variety of quantum phase transitions (*9*, *10*) and out-of-equilibrium spin dynamics in optical lattices (*11–15*). Ultracold polar molecules offer the advantage of stronger, tunable dipolar interactions (*16–21*). With electric field ($\boldsymbol{E}$)-tunable interactions between pseudo-spins encoded in rotational states, effective spin and spin-motion models can be implemented with high controllability (*22*) if two-body loss is suppressed by utilizing either strong confinement into lower dimensions (*23*) or collisional shielding with microwaves (*24–30*) or static electric fields (*31–35*). The extended lifetime of these itinerant molecules enables the combination of motion and strong dipolar interactions to study exotic spin-motion phenomena. Here, we report the realization of a generalized *t-J* model using ultracold dipolar molecules and the exploration of the system's out-of-equilibrium dynamics with independent control over both the motion and the interaction anisotropy $\chi = J_Z - J_\perp$ between spin-exchange ($J_\perp$) and Ising ($J_Z$) couplings in an XXZ spin model.

## Dipolar interactions between molecules in optical lattices

In our experiment, we prepared degenerate gases of $^{40}$K and $^{87}$Rb in a crossed optical dipole trap and loaded these atoms into the ground band of a 3D lattice with spacing $a_y = 540$ nm in the vertical direction and $a_x = a_z = 532$ nm in the horizontal directions. The atoms were then converted to fermionic KRb molecules (*36*) in their rovibrational ground state $|0\rangle$, where $|N\rangle$ is the rotational state with $\boldsymbol{E}$-dressed quantum number $N$ and projection $m_N = 0$ onto the quantization axis set by $\boldsymbol{E}$. The highest observed average molecular filling fraction was about 13%. For the spin measurements throughout this work, we used the two lowest rotational states of the molecule $|0\rangle$ and $|1\rangle$ as our spin-1/2 manifold, referred to as $|\downarrow\rangle$ and $|\uparrow\rangle$ respectively.

Heteronuclear molecules intrinsically have field-tunable dipolar interactions. The rotational states have induced dipole moments $d_\downarrow = \langle\downarrow|d_y|\downarrow\rangle$ and $d_\uparrow = \langle\uparrow|d_y|\uparrow\rangle$ and transition dipole moment $d_{\downarrow\uparrow} = \langle\downarrow|d_y|\uparrow\rangle$, where $d_y$ is the dipole operator along $\hat{y}$, the quantization axis set by $\boldsymbol{E}$. These dipole moments depend on $|\boldsymbol{E}|$, with $d_\downarrow \approx d_\uparrow \approx 0$ at $|\boldsymbol{E}| \approx 0$, and $|d_\downarrow|, |d_\uparrow| > 0$ at $|\boldsymbol{E}| > 0$, as other rotational states get mixed into the field-dressed state $|N\rangle$. The dipolar interactions and the motional confinement provided by optical lattices are illustrated in Fig. 1, A and B.

Dipolar interactions are described by

$$\hat{H}_{\text{int}} = h\int d^3\boldsymbol{r}\int d^3\boldsymbol{r}'\,\frac{[1-3\cos^2(\theta)]a_x^3}{2|\boldsymbol{r}-\boldsymbol{r}'|^3}\Bigg[\frac{J_\perp}{2}[\hat{s}^+(\boldsymbol{r})\hat{s}^-(\boldsymbol{r}') + \hat{s}^+(\boldsymbol{r}')\hat{s}^-(\boldsymbol{r})] + J_Z\hat{s}^Z(\boldsymbol{r})\hat{s}^Z(\boldsymbol{r}') + V\hat{n}(\boldsymbol{r})\hat{n}(\boldsymbol{r}') + W[\hat{n}(\boldsymbol{r})\hat{s}^Z(\boldsymbol{r}') + \hat{s}^Z(\boldsymbol{r})\hat{n}(\boldsymbol{r}')]\Bigg] \tag{1}$$

where $\hat{\psi}_\sigma(\boldsymbol{r})$ is the fermionic field operator at position $\boldsymbol{r}$ with spin σ (*7*, *8*, *36*), $\hat{n}(\mathbf{r}) = \sum_\sigma \hat{\psi}_\sigma^\dagger(\boldsymbol{r})\,\hat{\psi}_\sigma(\boldsymbol{r})$ is the density, $\hat{s}^Z(\boldsymbol{r}) = [\hat{\psi}_\uparrow^\dagger(\boldsymbol{r})\hat{\psi}_\uparrow(\boldsymbol{r}) - \hat{\psi}_\downarrow^\dagger(\boldsymbol{r})\hat{\psi}_\downarrow(\boldsymbol{r})]/2$ is the local magnetization, and $\hat{s}^+(\boldsymbol{r}) = \hat{\psi}_\uparrow^\dagger(\boldsymbol{r})\hat{\psi}_\downarrow(\boldsymbol{r})$ ($\hat{s}^-(\boldsymbol{r}) = [\hat{s}^+(\boldsymbol{r})]^\dagger$) raises (lowers) the spin. $\theta$ is the angle between $\boldsymbol{E}$ and $\boldsymbol{r}-\boldsymbol{r}'$. $hJ_\perp = 2d_{\downarrow\uparrow}^2/(4\pi\epsilon_0 a_x^3)$ is the spin-exchange interaction, $hJ_Z = (d_\downarrow - d_\uparrow)^2/(4\pi\epsilon_0 a_x^3)$ is the Ising interaction arising from differential dipole moments of the two states, $hV = (d_\downarrow + d_\uparrow)^2/(16\pi\epsilon_0 a_x^3)$ is the density-density interaction due to average dipole moments, and $hW = (d_\uparrow^2 - d_\downarrow^2)/(8\pi\epsilon_0 a_x^3)$ is the residual spin-density interaction arising from the cross-term of differential and average dipole moments. The relationship between dipole moments and $J_\perp, J_Z, V$, and $W$ is illustrated in Fig. 1C. The full Hamiltonian is $\hat{H} = \hat{H}_{\text{sp}} + \hat{H}_{\text{int}}$ (*36*) where $\hat{H}_{\text{sp}} = \int d^3\boldsymbol{r}\sum_\sigma \hat{\psi}_\sigma^\dagger(\boldsymbol{r})H_0\,\hat{\psi}_\sigma(\boldsymbol{r})$ and $H_0 = -\frac{\hbar^2}{2m}\boldsymbol{\nabla}^2 + \sum_{\alpha=x,y,z} V_{\text{L},\alpha}\sin^2(\frac{\pi r_\alpha}{a_\alpha})$ describe the kinetic and lattice potential energies, respectively (for simplicity, we neglect an external confining potential). Here, $\hbar = h/2\pi$ is the reduced Planck constant, $m$ is the mass of $^{40}K^{87}Rb$, and $V_{\text{L},\alpha}$ is lattice depth in the $\alpha \in (x, y, z)$ direction.

In a deep 3D optical lattice, molecules cannot move, and the Hamiltonian becomes a spin model of frozen dipoles. In contrast, in the absence of horizontal lattices ($V_{\text{L},x} = V_{\text{L},z} = 0$), molecules are fully itinerant, so the Hamiltonian describes quasi-2D dipolar collisions where the motion is confined to two dimensions, but the interaction potential remains three dimensional (*37*). Between these two extremes, both motion and interactions become discretized on a lattice, and the full Hamiltonian reduces to a generalized *t*-*J* Hamiltonian, i.e. the $t$-$J$-$V$-$W$ model (*7, 8*)

$$\hat{H}_{tJVW} = -h \sum_{\langle i,j \rangle,\sigma} t_{ij}\left[\hat{c}_{i\sigma}^{\dagger}\hat{c}_{j\sigma} + h.c.\right] + h \sum_{i \neq j} \frac{V_{ij}}{2} [J_{\perp}(\hat{\boldsymbol{s}}_i \cdot \hat{\boldsymbol{s}}_j) + \chi\left(\hat{s}_i^Z \hat{s}_j^Z\right) + V \hat{n}_i \hat{n}_j + W\left(\hat{n}_i \hat{s}_j^Z + \hat{n}_j \hat{s}_i^Z\right)] + hU \sum_i \hat{n}_{i\downarrow}\hat{n}_{i\uparrow} \quad (2)$$

where $\hat{c}_{i\sigma}^{\dagger}$ is the fermionic creation operator at lattice site $i$ for spin σ, and we have $\hat{n}_{i\sigma} = \hat{c}_{i\sigma}^{\dagger}\hat{c}_{i\sigma}$, total site occupation $\hat{n}_i = \sum_{\sigma} \hat{n}_{i\sigma}$, and magnetization $\hat{s}_i^Z = (\hat{c}_{i\uparrow}^{\dagger}\hat{c}_{i\uparrow} - \hat{c}_{i\downarrow}^{\dagger}\hat{c}_{i\downarrow})/2$. $t_{ij}$ characterizes the hopping rate of molecules between neighboring sites $i$ and $j$, denoted $\langle i, j \rangle$. $V_{ij}$ captures the geometric $(1 - 3\cos^2\theta)a_x^3/r^3$ component of the interactions. Strong kHz-scale on-site interaction $U$, rapid on-site two-body loss, and low filling fractions prohibit double occupancy of lattice sites and suppress superexchange interactions. The spin interaction anisotropy $\chi$ parameterizes rich spin dynamics that were studied at the mean field level in (*22*) and produces one-axis twisting that can be used to generate entangled states (*38*, *39*). The $\boldsymbol{E}$-tunability of the interactions for molecules on neighboring lattice sites in the $\hat{x}$-$\hat{z}$ plane is shown in Fig. 1D. We note that $V$ is spin-independent and small compared to spin-spin interactions and thus not expected to appreciably affect the spin dynamics observed here (*36*).

**Coherent spin dynamics probed with Ramsey spectroscopy**

To explore the out-of-equilibrium coherent dynamics of the molecules in different interaction regimes and geometries of the generalized *t-J* model, we measured the Ramsey contrast decay (Fig. 2A). First, an applied microwave pulse prepared an equal superposition of the $|\downarrow\rangle$ and $|\uparrow\rangle$ states. In a Bloch sphere picture, the collective Bloch vector was prepared along $\hat{X}$ with maximal length $N_{\text{mol}}/2$ for total molecule number $N_{\text{mol}}$. The system was allowed to evolve for a variable time $T$ before a microwave pulse of area $\pi/2$ about variable axis $\hat{R} = \cos\varphi\,\hat{X} + \sin\varphi\,\hat{Y}$ was applied. The number of molecules in each spin state was then measured (*36*). By repeating the measurement several times while varying the phase $\varphi$ of the second pulse and taking the standard deviation of the fraction in the excited state, the contrast, or the equatorial length of the Bloch vector, $C = \frac{2}{N_{\mathrm{mol}}}\sqrt{\langle\sum_i \hat{s}_i^X\rangle^2 + \langle\sum_i \hat{s}_i^Y\rangle^2}$ at wait time $T$ was extracted (*36*). Since $\langle\sum_i \hat{s}_i^Z\rangle = 0$ is conserved during the dynamics, $C$ describes the magnetization of the sample. During $T$, a KDD pulse sequence (*40*) suppresses single-particle dephasing by frequently swapping the populations in the $|\downarrow\rangle$ and $|\uparrow\rangle$ states (*36*). Dynamical decoupling also removes the $W$ term in the $t$-$J$-$V$-$W$ Hamiltonian. The contrast evolution was then fit to a stretched exponential via $C(T) = e^{-(\Gamma T)^{\nu}}$ where $\Gamma$ is the dephasing rate and $\nu$ is the stretching parameter. We observe $\nu < 1$ for all parameters analyzed in this work, describing sub-exponential decay, possibly associated with glassy dynamics (*41*) and number loss (*36*).

To examine the effect of field-tunable interactions on spin coherence, we first freeze the motional degree of freedom by confining the molecules in a deep 3D optical lattice ($t_{ij} \approx 0$ in all directions). At zero electric field, $d_\downarrow \approx d_\uparrow \approx 0$, molecules interact predominately via spin-exchange, realizing the XY model $\hat{H}_{XY} = h\sum_{i\neq j}\frac{V_{ij}}{2}J_\perp(\hat{s}_i^X\hat{s}_j^X + \hat{s}_i^Y\hat{s}_j^Y)$. The coherent spin dynamics in this regime at low fillings have been studied (*16*, *17*, *42*) revealing density-dependent Ramsey contrast decay from long-range interactions, with periodic revivals in contrast arising from spin exchange between nearest neighbors. Turning on static $\boldsymbol{E}$ to add Ising interactions realizes the XXZ model $\hat{H}_{XXZ} = h\sum_{i\neq j}\frac{V_{ij}}{2}[J_\perp(\hat{\boldsymbol{s}}_i\cdot\hat{\boldsymbol{s}}_j) + \chi\,(\hat{s}_i^Z\hat{s}_j^Z)]$. Previous works have studied the coherent dynamics of this model in the presence of strong disorder utilizing Floquet engineering on platforms such as nitrogen-vacancy centers in diamond (*43*) and Rydberg atoms (*44*). Here we utilize polar molecules to probe these lattice spin models beyond the pure spin-exchange regime. At the Heisenberg point ($\chi = 0$ at $|\boldsymbol{E}|$ = 6.5 kV/cm), the Hamiltonian reduces to the isotropic XXX model $\hat{H}_{XXX} = h\sum_{i\neq j}\frac{V_{ij}}{2}J_{\perp}(\hat{\boldsymbol{s}}_i \cdot \hat{\boldsymbol{s}}_j)$, where the interactions are independent of spin orientation, making all points on the collective Bloch sphere eigenstates of the Hamiltonian. As such, there should be no interaction-induced dephasing. Figure 2B shows contrast decay traces and fits to stretched exponentials for average two-dimensional densities of roughly $1.5 \times 10^7$ cm$^{-2}$ at two different values of $\boldsymbol{E}$ where $\chi = 0$ (blue circles) and $\chi = 102$ Hz (orange squares, $|\boldsymbol{E}|$ = 12.72 kV/cm). We observed an order of magnitude slower contrast decay at the Heisenberg point than at $\chi = 102$ Hz.

If interactions limit the coherence time, higher densities should lead to faster contrast decay. To extract the effect of many-body processes, the contrast decay measurement is repeated for different initial average 2D densities $n$ (*36*). The contrast decay rates for each $n$ are fit to a linear function $\Gamma(n) = \kappa n + \Gamma_0$, where $\Gamma_0$ is the single-particle dephasing rate after dynamical decoupling and $\kappa$ is the many-body dephasing rate. This density-normalization procedure is shown in Fig. 2C for $\chi = 0$ (blue circles) and $\chi = 102$ Hz (orange squares). Prominently, in the $\chi = 0$ case, we measured $\kappa = 0.07(7) \times 10^{-6}$ cm$^2$ s$^{-1}$, consistent with no density-dependent contrast decay, as expected in the XXX model with a single-body dephasing rate, determined from the linear slope offset, of $\Gamma_0 = 4.3(7)$ s$^{-1}$. By contrast, for $\chi = 102$ Hz, we observed contrast decay rates that depend strongly on density, with $\kappa = 2.6(4) \times 10^{-6}$ cm$^2$ s$^{-1}$ and $\Gamma_0 = 6(3)$ s$^{-1}$.

**Interaction-limited decoherence dynamics**

Repeating the measurement at several values of $|\boldsymbol{E}|$ in the pinned configuration, we observed a roughly linear dependence of $\kappa$ on $|\chi|$, with a slight dependence on the sign of $\chi$ (Fig. 3A). $\kappa$ depends approximately linearly on $|\chi|$ because dephasing in our lattice originates from couplings with strength proportional to $\chi$ within local clusters of molecules. We attribute the observed slower decay in the $J_{\perp}$-dominated ($\chi < 0$) regime to periodic self-rephasing from spin-exchange within local clusters that is suppressed in the $J_Z$-dominated ($\chi > 0$) regime (*36*). We also model the contrast dynamics using a moving-average cluster expansion (MACE) (*36*, *42*). In this approximation, the dynamics for a given molecule are solved exactly in the presence of its $M$ most strongly coupled neighbors. The global contrast dynamics are then obtained by averaging these "cluster" dynamics for every molecule in the system, with reasonable convergence being obtained for clusters as small as $M = 6$. This convergence for small particle number is an indicator of the local nature of the spin dynamics under anisotropic dipolar interactions in 3D, whereas the spin dynamics for isotropic interactions can depend collectively on all particles in the system in certain geometries (*45–47*). The results of the MACE simulation (gray trace in Fig. 3A) show excellent agreement between theory and experiment and confirm that pinned dipolar spin dynamics are well-described by the XXZ model.

Next, we observed modifications to the many-body dephasing rates from collisions of fully itinerant molecules. To probe the interplay of motion with spin-dependent dipolar interactions, we confined the molecules to 2D layers formed by a vertical lattice of $65\,\mathrm{E_r}$ depth, where $\mathrm{E_r} \approx h \times 1.4$ kHz is the photon recoil energy for KRb. In these experiments, without any horizontal lattice potential ($V_{\mathrm{L},x} = V_{\mathrm{L},z} = 0$), radial harmonic confinement is provided predominantly by a crossed optical dipole trap, which is also maintained for all other experiments. We then extract the density-normalized dephasing rate $\kappa$ as a function of $\chi$ as in the pinned molecule case, (green squares in Fig. 3B). In prior work (*22*), the short time mean-field dynamics were well-described by an XXZ spin model with spins pinned in a harmonic oscillator mode space lattice (*36*). At longer times relevant for the current measurements, the thermal occupation of our sample enables molecules to delocalize in mode space caused by mode-changing collisions, invalidating the pinned mode picture. Although collisional decoherence was previously observed in ultracold molecules (*22*), the dependence on dipolar interactions was not completely understood, and modeling dipolar collisional dephasing of coherent superpositions has been challenging.

We systematically measured that $\kappa$ depends more strongly on $|\chi|$ than in the pinned case, and we present a scattering approach to study the $V_{\mathrm{L},x} = V_{\mathrm{L},z} = 0$ limit of the molecular Hamiltonian to explain our observations. We adopt two methods of theoretical analysis (*36*): collisional Monte Carlo simulations and a simplified analytic model. Data obtained from Monte Carlo simulations (gray trace in Fig. 3B) show agreement with the experimental data, apart from experimentally measured negative values of $\kappa$. We attribute these negative values to the preferential two-body loss of decohered molecules (*36*). The agreement between theory and experiment suggests that our theoretical model captures the relevant physics from collisional dephasing of molecular superpositions.

The simplified analytic model provides a more intuitive understanding of the observed stronger, approximately cubic, dependence of $\kappa$ on $|\chi|$. With the molecular spins initially prepared in the state $|X\rangle = (|\uparrow\uparrow\rangle + |\downarrow\downarrow\rangle + |\uparrow\downarrow\rangle + |\downarrow\uparrow\rangle)/2$, their first collision would only involve states in the symmetric spin sector: $|\downarrow\downarrow\rangle$, $(|\uparrow\downarrow\rangle + |\downarrow\uparrow\rangle)/\sqrt{2}$, and $|\uparrow\uparrow\rangle$, which each accrue a scattering phase shift. Collisions in this sector are predominantly elastic, as Fermi statistics suppresses close contact of the molecules, implying that long-ranged dipolar interactions dominate these phase shifts. We then expect that completed collisions are slow compared to the dynamical decoupling pulses that rapidly flip molecules between $|\downarrow\rangle$ and $|\uparrow\rangle$. The result is that the aligned ($\updownarrow$) $|\downarrow\downarrow\rangle$ and $|\uparrow\uparrow\rangle$ spin states will both incur the same scattering phase shift $\delta_{\updownarrow}$ proportional to dipole length (*48*): $a_D^{\updownarrow} \propto d_{\downarrow}^2 + d_{\uparrow}^2 \propto V + \frac{J_{\perp}}{4} + \frac{\chi}{4}$. The remaining anti-aligned triplet state ($\leftrightarrow$) will, in general, develop a different phase shift $\delta_{\leftrightarrow}$ proportional to the dipole length $a_D^{\leftrightarrow} \propto d_{\downarrow\uparrow}^2 + d_\downarrow d_\uparrow \propto V + \frac{J_\perp}{4} - \frac{\chi}{4}$. Each single collision thus causes the scattered molecules to decohere, with a change in contrast $\Delta C = 2\sin^2(\delta_{\leftrightarrow} - \delta_{\updownarrow}) \propto \chi^2$ for small $\delta_{\leftrightarrow} - \delta_{\updownarrow} \propto \chi$ (see inset of Fig. 3B for schematic). The total dephasing rate is approximated by the product of the change in contrast for each collision, and the elastic collision rate: $\kappa \propto \Delta C \beta_{el}$. For 2D motion, ultracold identical spin-polarized fermionic molecules scatter elastically at a rate proportional to their threshold cross section $\sigma \approx 8\pi k a_D^2$ where $k$ is the relative momentum of the colliding pair (*23*, *37*, *48*). Molecules in superpositions should also elastically scatter from both aligned and anti-aligned channels, so that $\beta_{\mathrm{el}} = (\beta_{\updownarrow} + \beta_{\leftrightarrow}) / 2 = 4\pi\hbar k^2 \left[ \left(a_D^{\updownarrow}\right)^2 + (a_D^{\leftrightarrow})^2 \right] / \mu$ where $\mu$ is the reduced mass of the pair. Evaluating the product for $\kappa$ captures the dominant $\kappa \propto |\chi|^3$ scaling seen to arise in the Monte Carlo simulations (*36*). Notably, we find that this scaling is sensitive to the trap frequency of the tight confining axis, and is thus relevant to our current experiment but not universal.

### Spin dynamics of the generalized *t-J* model

We next explored the more complex generalized *t-J* model regime by tuning tunneling within 2D layers. Experimentally, we studied the effect of motion on contrast by measuring $\kappa$ of molecules confined in a deep vertical lattice with variable corrugation in the $\hat{x}$ and $\hat{z}$ directions, for different values of $\chi$. The measured values of $\kappa$ as a function of tunneling rate $t$ in both the $\hat{x}$ and $\hat{z}$ directions are plotted in Fig. 4A, with black circles, blue squares and orange triangles representing $\chi = -205$ Hz, $\chi = 0$ Hz, and $\chi = 102$ Hz ($J_\perp = 211$ Hz, $J_\perp = 133$ Hz, and $J_\perp = 75$ Hz), respectively. For $\chi = 0$ Hz, we observed no density-dependent contrast decay (that is, $\kappa$ is consistent with 0) over the entire range of $t$ explored, as intuitively expected at the Heisenberg point. For $\chi = 102$ Hz, we saw a smooth transition between the two motional extremes, with $\kappa$ gradually decreasing for increasing $t$. For $\chi = -205$ Hz we observed stronger decoherence in general with a peak in decoherence rate around $t = 70$ Hz. Similar to the fully itinerant case, and as a result of the nonlinear dependence of collisional dephasing on $|\chi|$, it is not surprising that the behavior for $\chi = -205$ Hz can be qualitatively different than for $\chi = 102$ Hz.

To understand the observed peak when $\chi = -205$ Hz, we first examined the rise in $\kappa$ as the lattice depth was reduced from the pinned molecule case. Intuitively, increasing tunneling will couple the internal and external degrees of freedom of the molecules, leading to increased dephasing. This mechanism is well captured by simulations with an extended MACE (EMACE) (*36*), which adds motion to the standard MACE, and is shown as a light gray (orange) band for $\chi = -205$ Hz ($\chi = 102$ Hz) in Fig. 4A.

We next examined the enhanced contrast decay as the lattice depth was increased from zero. From a band structure perspective, increasing the lattice depth shrinks the width of each motional band, increasing its density of states and allowing more channels for mode-changing collisions, generally leading to faster spin decoherence ((*36*), Fig. 4B). From a scattering perspective, in the lattice, molecules move slower because of larger effective mass (*49*), increasing the phase shifts accrued in each collision, enhancing contrast decay. In addition, molecular losses from inelastic collisions increase as the lattice depth is reduced (*36*), and the losses are largest for molecules that have already decohered, suppressing contrast decay in the remaining sample (*36*). While the EMACE simulation qualitatively captures the peak structure, quantitative differences arise owing to neglected molecular loss and limited cluster sizes. To further understand the role of mode-changing collisions and molecular loss in shallow lattices, a two-body simulation based on Hamiltonian trotterization (*36*) is included for $t > 50$ Hz as a black (orange) solid line for $\chi =$ $-205$ Hz ($\chi = 102$ Hz) in Fig. 4A, which agrees well with the experimental data with an empirical rescaling of 0.75.

Finally, we studied the role of motional dimensionality on the spin dynamics by measuring the dependence of $\kappa$ on $t$ for molecules allowed to tunnel in only one direction (Fig. 4C). Whereas the $\chi = 0$ Hz and $\chi = 102$ Hz dynamics are almost identical to the 2D case, noticeably, the peak in contrast decay around $t \approx 70$ Hz for $\chi = -205$ Hz disappears when the kinetic dimensions are reduced from two to one. In one dimension (1D), we instead only observe a rise in $\kappa$ at very shallow lattices ($t \gtrsim 200$ Hz). We attribute the different $\chi = -205$ Hz dynamics in the 1D case to the interplay of the confining harmonic trap with the lattice. For tunneling in only one direction, the combined potential leads to quasi-localization of the molecules owing to potential energy differences between neighboring lattice sites (*50*, *51*), schematically shown in Figure 4E . In the 1D configuration, most of the molecules are localized along all three axes of motion (*36*), leading to dynamics qualitatively similar to the motionless case as the lattice depth is lowered. In contrast, for the 2D case, the additional tunneling dimension allows these particles to delocalize within a quasi-1D azimuthal ring (Fig. 4D).

An EMACE simulation (gray (orange) band for $\chi = -205$ Hz ($\chi = 102$ Hz)) in Fig. 4C shows favorable agreement with experimental values of $\kappa$ as $t$ is varied in one direction. Note that the number of available tunneling sites is smaller in one dimension than in two, enabling more controlled EMACE simulations for shallow lattices in the 1D case than in the 2D case (*36*). As shown in both the experimental data and EMACE simulations, the 1D dynamics remain relatively flat between the pinned and fully itinerant cases, consistent with the quasi-localization picture. We attribute the slight disagreement between theory and experiment at very shallow lattices ($t \gtrsim 200$

Hz) for $\chi = 102$ Hz to molecular loss not included in EMACE and to the breakdown of the single-band tight-binding model.

## Discussion and outlook

Here, we presented a systematic study of the out-of-equilibrium dynamics of a generalized *t-J* system of interacting polar molecules. By increasing molecular filling fractions and using single-site detection (*6*, *17*) or spectroscopic signatures (*52*, *53*), it should be possible to study equilibrium states of these dipolar Hamiltonians. Using similar experimental techniques, a future out-of-equilibrium investigation could probe spin diffusion in lattice-confined polar molecules, potentially observing many-body localized states (*54*). Additionally, with an isolated single layer (*18*) and improved detection, the generalized *t-J* model is predicted to dynamically generate highly entangled spin squeezed states with applications to precision measurement (*38*, *39*). Recently, we demonstrated Floquet engineering of XYZ spin models, including a two-axis twisting Hamiltonian (*55*) which could enable more scalable spin squeezing than that produced with native XXZ models. Ultracold molecules, whose dipolar interactions can be tuned independently of motion, have thus emerged as a versatile platform to study a broad range of itinerant spin problems in many-body physics.

**Acknowledgements:** We thank Nelson Darkwah Oppong and Alexey V. Gorshkov for a careful review of this manuscript and for providing useful comments. We acknowledge experimental contributions from Jun-Ru Li and Jacob S. Higgins, and helpful discussions with Philip J. D. Crowley and Norman Y. Yao.

**Funding:** This material is based upon work supported by the National Science Foundation grant no. QLCI OMA-2016244. Additional support is acknowledged from the US Department of Energy, Office of Science, National Quantum Information Science Research Centers, Quantum Systems Accelerator, ARO and AFOSR MURIs, the JILA Physics Frontier Center grant no. PHY-2317149, the National Science Foundation grant no. PHY-2110327, the ARO single investigator Award No. W911NF-24-1-0128, and the National Institute of Standards and Technology. A.N.C acknowledges support from the National Science Foundation Graduate Research Fellowship under grant no. DGE 2040434. C.M. acknowledges support from the Department of Defense through the NDSEG Graduate Fellowship. S.R.M. acknowledges support from the National Science Foundation under grant no. QLCI OMA-2120757. K.P.Z. acknowledges support from the Austrian Science Fund (FWF) under grant no. W1259-N27. R.R.W.W acknowledges partial support from the NSF through a grant from ITAMP at Harvard University.

**Author Contributions:** A.N.C, H.H., C.M., J.L., K.P.Z, and J.Y. conducted the experiments and analyzed the data. D.W., S.R.M., R.R.W.W., J.L.B, and A.M.R developed theoretical models. All authors contributed to interpreting results and writing the manuscript.

**Competing Interests:** The authors declare no competing interests.

**Data and Materials Availability:** All data needed to evaluate the conclusions in this paper are available through Zenodo (*56*).

## Supplementary Materials

Materials and Methods

Supplementary Text

Figures S1-S16

Table S1

References (*57–73*)

**Figures:**

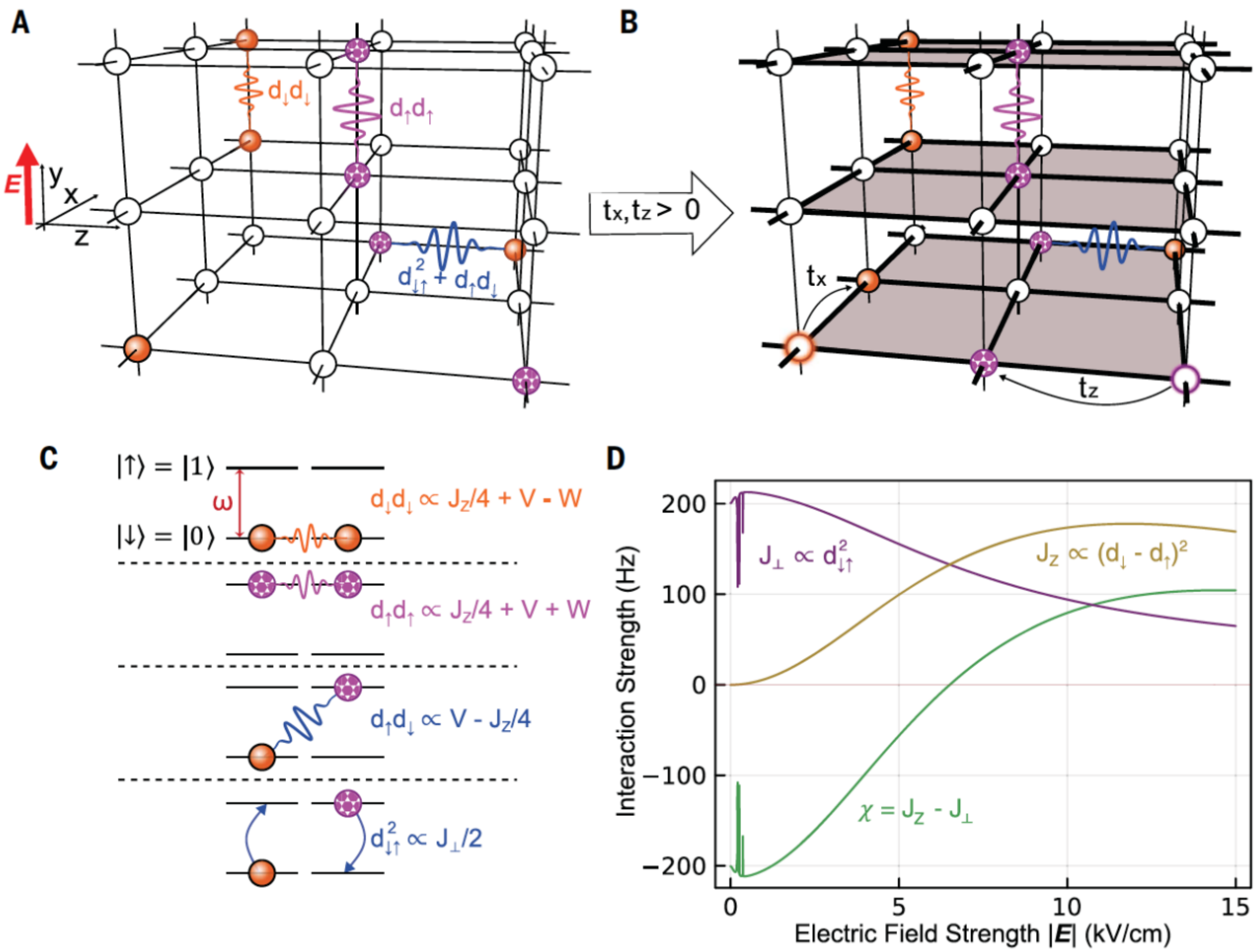

**Fig. 1: Field-tunable dipolar interactions between lattice-confined molecules. (A)** Molecules sparsely occupy a deep 3D optical lattice. Sites are shaded white if unoccupied, pink patterned if in the |↑⟩ state, and solid orange if in the |↓⟩ state. Molecules interact via induced dipole moments and transition dipole moments represented by squiggly lines between lattice sites. **(B)** Lowering

the lattice depth in the horizontal directions allows tunneling between sites within layers, represented by the black arrows $t_x, t_z$. For most of our experiments $t_x = t_z$. **(C)** Molecular interactions arising from dipole moments can be rewritten in terms of $J_\perp$, $J_Z, V$, and $W$, the spin-exchange, Ising, density-density, and spin-density interactions, respectively, which are used in a spin-basis Hamiltonian. The interactions set by $d_\downarrow^2, d_\uparrow^2, d_\uparrow d_\downarrow,$ and $d_{\downarrow\uparrow}^2$, and their associated Hamiltonian terms, are drawn schematically from top to bottom. The two lowest rotational states of the molecules, $|\downarrow\rangle$ and $|\uparrow\rangle$, are split by microwave-frequencies $\omega$ (red arrow between the spin levels in top row) between 2.2 GHz and 4.2 GHz depending on electric field strength ($|\boldsymbol{E}|$). **(D)** Calculated dipolar interaction strengths for KRb molecules separated by 532 nm perpendicular to the dipole orientations as a function of $|\boldsymbol{E}|$. Purple represents the spin-exchange interaction $J_\perp$ arising from the transition dipole moment between a $|\downarrow\rangle$ molecule and a $|\uparrow\rangle$ molecule which decreases with increasing electric field magnitude, gold represents the Ising interaction $J_Z$ from induced dipole moments which increases with increasing field strength, and green represents the interaction-type anisotropy $\chi = J_Z - J_\perp$ which crosses zero around 6.5 kV/cm. Hyperfine structure of the molecules produces interaction strengths that change dramatically with small changes in $|\boldsymbol{E}|$ for small electric fields less than 1 kV/cm.

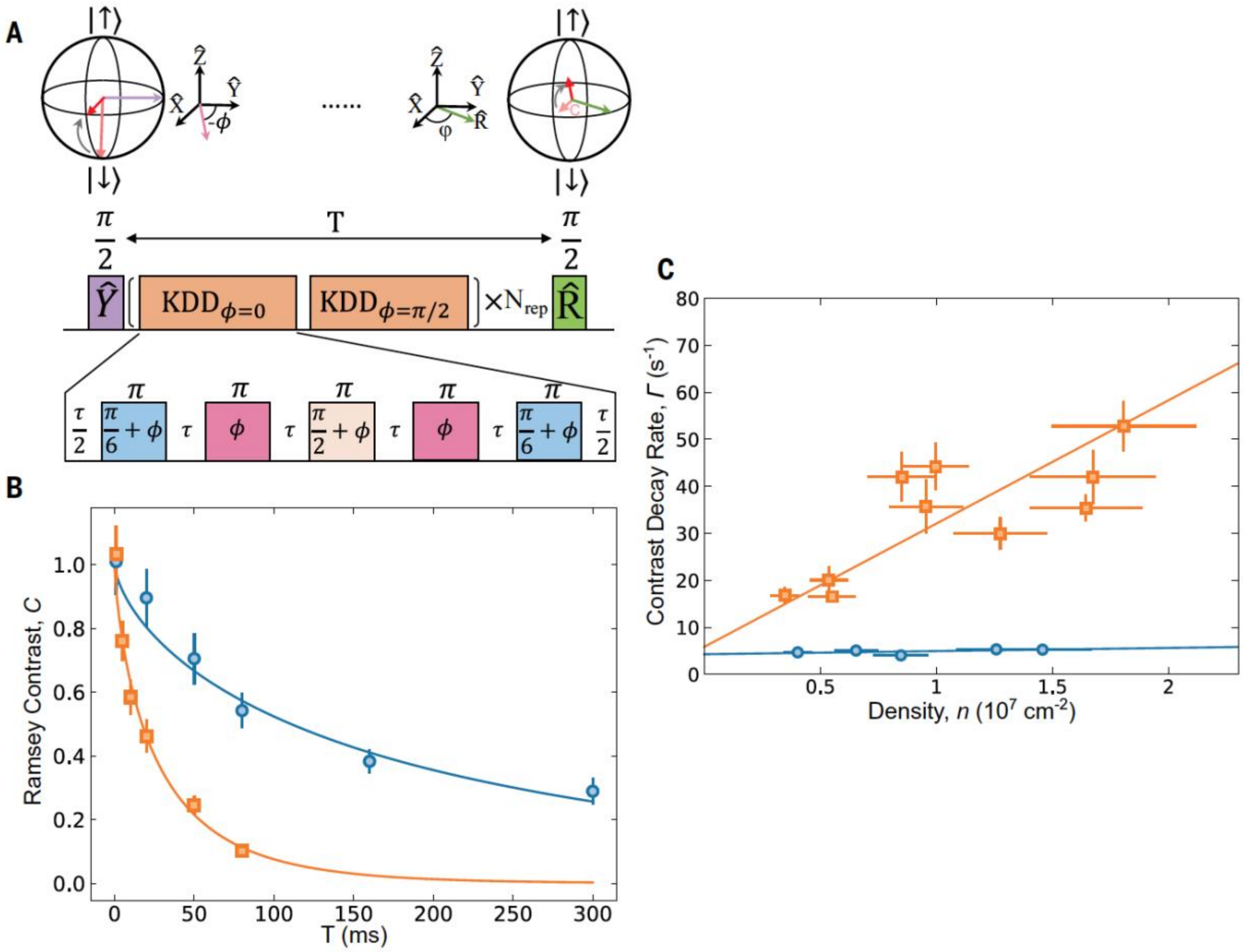

**Fig. 2: Dynamical magnetization of interacting pinned molecules. (A)** Ramsey spectroscopy is used to measure the dynamical magnetization of the molecules. The top row shows the orientation of the collective spin in the Bloch sphere representation, and the middle and bottom rows show the pulse sequence. A $\pi/2$ pulse about the $\hat{Y}$ axis prepares the molecules in a coherent superposition of $1/\sqrt{2}(|\downarrow\rangle + |\uparrow\rangle)$ . A KDD pulse sequence (shown as the bottom row) removes single-particle dephasing and is repeated a variable number of times $N_{rep}$ to extend the total interrogation time to $T$. A final $\pi/2$ pulse about variable axis $\hat{R}$ reads out the projection of the Bloch vector orthogonal to $\hat{R}$. By repeating the measurements varying $\hat{R}$, the equatorial length of the Bloch vector, which is the Ramsey contrast (dynamical magnetization) $C$, at time $T$ is extracted. **(B)** Measured contrast decay $C(T)$ for $\chi = 0$ (blue circles) and $\chi = 102$ Hz (orange squares) for initial 2D average densities ($n$) of roughly $1.5 \times 10^7$ cm$^{-2}$ of molecules confined in a deep 3D optical lattice. Solid lines are stretched exponential fits to the experimental data. Error bars are 1 s.d. from bootstrapping *(36)*. **c)** Extracted Ramsey contrast decay rates versus initial density for $\chi = 0$ (blue circles) and $\chi = 102$ Hz (orange squares). Solid lines are linear fits to the data whose slopes measure density-dependent decoherence rate $\kappa$. Error bars are 1 s.e. from fits (stretched exponential fit for contrast decay rates, one-body loss for densities).

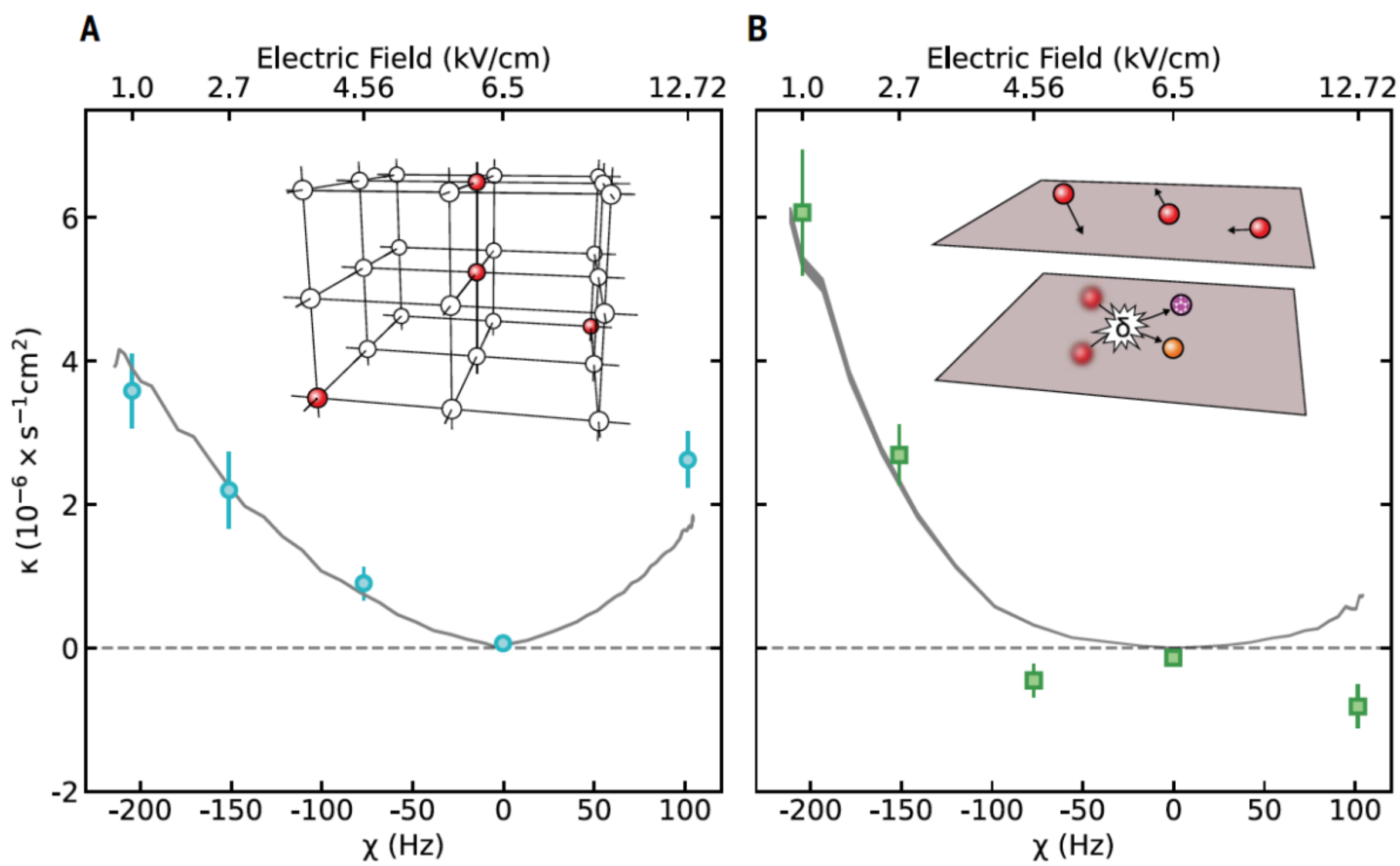

**Fig. 3: Field-tunable density-dependent decoherence rates. (A)** Density-normalized contrast decay rates $\kappa$ for pinned molecules tuned with interaction anisotropy $\chi$. Blue circles are experimental data extracted from slopes in Fig. 2C. Error bars are 1 s.e. from linear fits. Gray line is a MACE simulation (36) with the same density-normalization procedure as experimental data with no scaling. Inset: 3D lattice with fully pinned molecules, with white circles representing

unoccupied sites and red circles denoting sites occupied by a molecule initially in $1/\sqrt{2}(|\downarrow\rangle + |\uparrow\rangle)$. **(B)** As in (A) but for molecules confined to 2D layers without horizontal corrugation. Green squares are experimental data. Error bars are 1 s.e. from linear fits. Gray shaded band is from Monte Carlo coherent collision simulations with no scaling *(36)*, with the band halfwidth representing 1 s.e. of linear fit. Inset: 2D geometry of the system, with molecules (red circles) free to move within 2D layers. In the bottom 2D layer, collisional dephasing occurs when a collision leads to a relative phase shift $\delta = \delta_{\leftrightarrow} - \delta_{\updownarrow}$, making the molecules no longer identical after the collision, illustrated as one molecule becoming pink and the other turning orange.

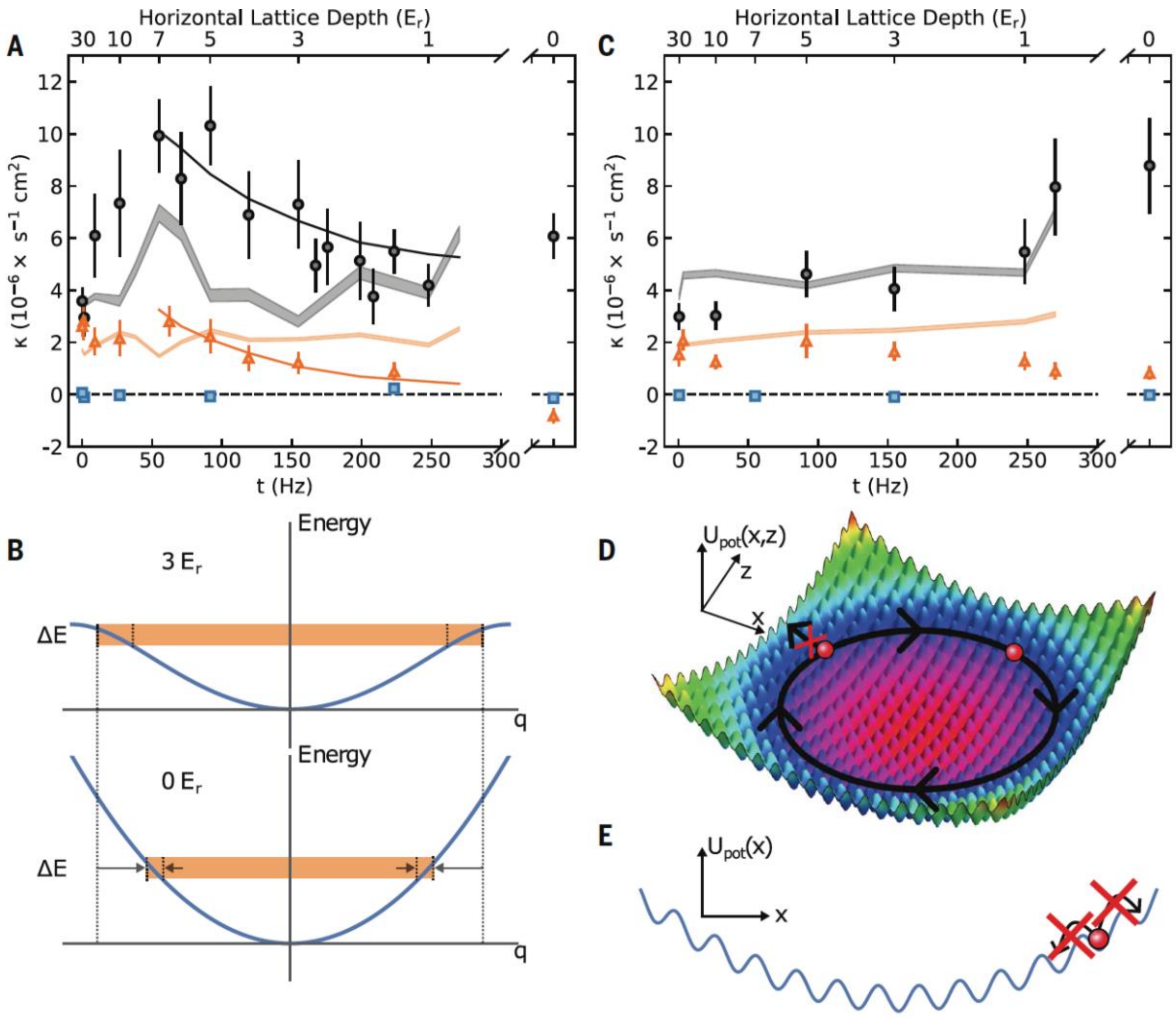

**Fig. 4: Tuning coherent *t-J* dynamics. (A)** Density-dependent contrast decay rates $\kappa$ versus tunneling rate $t$ in two directions. Black circles, blue squares, and orange triangles represent experimental data for $\chi = -205, 0,\ 102$ Hz respectively. Horizontal axis break occurs when transverse lattices are turned off (case of Fig. 3B). Error bars are 1 s.e. from linear fits. Light gray (orange) bands are EMACE simulations *(36)* for $\chi = -205$ Hz ($\chi = 102$ Hz) with no scaling, with the band halfwidth representing 1 s.e. of linear fit. Black (orange) solid lines shown for $t >$ $50$ Hz are two-body simulations *(36)* with a rescaling by 0.75 for $\chi = -205$ Hz ($\chi = 102$ Hz). **(B)** Top row: kinetic energy of two particles in a $3E_r$ lattice as a function of relative

quasimomentum $q$. Bottom row: kinetic energy in the absence of a lattice where $q = k$. Shaded orange regions show the range of momentum states within a given energy interval, showing the increased density of states in a shallow lattice compared to the no lattice case. **(C)** As in (A)**,** but for tunneling only in one direction. **(D)** 3D rendering of the single-molecule potential energy $U_{pot}$ in the $\hat{x}$ and $\hat{z}$ directions arising from a weak 2D optical lattice and a crossed optical dipole trap. Colors indicate equipotential surfaces. Because of the site-to-site energy shift, molecular tunneling along the radial directions is suppressed, and instead tunneling occurs along azimuthal rings. **(E)** Schematic of the localization of molecules within a potential landscape $U_{pot}$ produced by a weak 1D optical lattice in the $\hat{x}$ direction and a crossed optical dipole trap (equivalently, by taking a 1D slice of the 2D lattice potential shown in (D)) because of site-to-site energy shifts, when the motion is also frozen in the other two directions by deep optical lattices.

# Supplementary Materials for

## Observation of generalized *t-J* spin dynamics with tunable dipolar interactions

Annette N. Carroll, Henrik Hirzler, Calder Miller, David Wellnitz, Sean R. Muleady, Junyu Lin, Krzysztof P. Zamarski, Reuben R. W. Wang, John L. Bohn, Ana Maria Rey, Jun Ye

Corresponding authors: annette.carroll@colorado.edu (A.N.C) and ye@jila.colorado.edu (J.Y.)

**The PDF file includes:**

Materials and Methods
Supplementary Text
Figs. S1 to S16
Table S1

## Materials and Methods

Sample Preparation

We prepared samples of ultracold KRb molecules at each desired value of $\boldsymbol{E}$ following the procedure of (*22*) with the exception that the atomic transfer to the molecular state $|0\rangle = |N = 0, m_N = 0, m_{\mathrm{K}} = -4, m_{\mathrm{Rb}} = 1/2\rangle$ was performed in a 3D optical lattice instead of a 1D optical lattice, with the atoms predominately occupying the ground band in all directions. For the 2D measurements, molecules were magneto-associated and then optically transferred to the rovibrational ground state with a vertical lattice of depth 65 $\mathrm{E_r}$ and the transverse lattices depths set to 25 $\mathrm{E_r}$. Before performing Ramsey measurements, the intensities of the transverse lattices were adiabatically ramped to depths ranging from 0 $\mathrm{E_r}$ (the fully itinerant case of Fig. 3B) to 65 $\mathrm{E_r}$ (the pinned case of Fig. 3). For the 1D measurements, the molecules were instead produced in a vertical lattice and one horizontal lattice of 30 $\mathrm{E_r}$ and another horizontal lattice 25 $\mathrm{E_r}$ before that lattice was ramped to the target depth for the measurement. The reduction of depth in each tight direction is chosen to keep the temperature in 1D similar to that in 2D in the absence of corrugation in the weak direction(s) of approximately 300 nK. In all cases, the polarization of each horizontal lattice is tuned to match the ac polarizability of the $|1\rangle$ state to that of the $|0\rangle$ state so that changing lattice depths does not change single-particle dephasing rates (*57*). Additionally, a crossed optical dipole trap provides harmonic confinement in all lattice configurations, producing in net (including the weak contribution from the vertical lattice) radial trap frequencies of $(\omega_{x'}, \omega_{z'}) = 2\pi \times (36(4), 29(2))$ Hz where $\hat{x}' = (\hat{x} - \hat{z})/\sqrt{2}$ and $\hat{z}' = (\hat{x} + \hat{z})/\sqrt{2}$.

The experimental 2D average density is computed assuming uniform occupation of 19(1) layers, confirmed with layer-selection (see Methods of (*22*) for procedure), as $n = N_{\mathrm{mol}}/(19 \times 4\pi\sigma_{x'}\sigma_{z'})$, where $N_{\mathrm{mol}}$ is the total number of molecules imaged in both spin states. In general, our cloud size is roughly (16, 2.8, 20) $\mu$m, where we measured $\sigma_{z'}$ directly from in situ imaging and extract $\sigma_{x'}$ using the relative ratio between the trap frequencies in the $\hat{x}'$ and $\hat{z}'$ directions.

To vary the density uniformly across the sample, we follow the procedure of (*22*). In brief, we apply a microwave pulse of area $\theta$ to prepare a superposition of $|\downarrow\rangle$ and $|\uparrow\rangle$. We then apply an optical blast using the STIRAP beam resonant with $|\downarrow\rangle$ to remove the $|\downarrow\rangle$ fraction of the population. This procedure reduces the density by a factor $\sin^2\frac{\theta}{2}$. At $|\boldsymbol{E}| = 1$ kV/cm where the STIRAP beam also couples to $|\uparrow\rangle$, we shelve the $|\uparrow\rangle$ component in the $|2\rangle$ state before applying the cleaning optical pulse.

We perform state-resolved imaging of both spin states following (*18, 22*). In short, to image the molecules in $|0\rangle$, we reverse the STIRAP process and image K atoms in Feshbach molecules. To convert the measured number of atoms to the real number of molecules, we divide by the STIRAP efficiency (70-85% depending on $|\boldsymbol{E}|$) and the efficiency of imaging K in a Feshbach molecule (70%). By applying a $\pi$-pulse to transfer the population in $|1\rangle$ to $|0\rangle$ and repeating the imaging procedure, we also can detect the number of molecules in $|1\rangle$. Because of coupling of the STIRAP beam to $|1\rangle$ at $|\boldsymbol{E}| = 1$ kV/cm, to image both states at that field, we shelve $|1\rangle$ in $|2\rangle$ while imaging $|0\rangle$.

Lattice Depths and Tunneling Rates

We calibrated the lattice depths using parametric heating of Rb which measures the gap between the ground band and the second excited band (*58*) and then scaled these frequencies to those experienced by KRb molecules (approximately a factor of 1.1) (*59*). By calculating the expectation of the single particle Hamiltonian for two ground-band Wannier functions of adjacent sites for each lattice potential, the nearest-neighbor tunneling rate $t$ for a KRb molecule can be estimated ranging from 270 Hz at 0.5 $E_r$ to 0.01 Hz at 65 $E_r$. We report these calculated values as the primary x-axis in Fig. 4.

Contrast Measurement and Bootstrapping Error

At long times in non-zero electric fields, while molecule-molecule coherence can be preserved, fluctuations in the electric field that scramble phase information can lead to loss of molecule-microwave coherence. As such, we follow the procedure of (*18*) and (*22*). In short, rather than fit a Ramsey fringe, we calculate the number-normalized contrast from $\sigma_f(T, N_{\text{mol}})$, the standard deviation of spin fraction $f$ at time $T$ for total molecule number $N_{\text{mol}}$ over approximately 16 runs, as $C(T, N_{\text{mol}}) = 2\sqrt{2}\sqrt{\sigma_f^2(T, N_{\text{mol}}) - \sigma_0^2(N_{\text{mol}})}$ where $\sigma_0(N_{\text{mol}})$ is the non-zero contrast in the absence of molecule coherence arising from imaging noise. We determined $\sigma_0(N_{\text{mol}}) = 0.120(9) - 4(1) \times 10^{-6} N_{\text{mol}}$ by measuring the apparent contrast after 50 ms without dynamical decoupling for varied densities. To estimate the error in contrast at each time, we bootstrap the sampled set of fractions following the procedure of (*22*).

Role of Stretching Parameter

When there is density-dependent contrast decay, decay in deeper lattices tends to be best fit with smaller values of stretching parameter $\nu$ than in shallower lattice cases. A plot of the mean $\nu$ versus lattice depth for three interaction cases is provided as Fig. S1. This suggests that sub-exponential decay is dominated by glassy dynamics, as the loss is small in these deep lattice cases. This trend is persistent independent of interaction anisotropy strength $\chi$, except when $\chi = 0$ where it does not obviously correlate with lattice depth, further evidence that $\nu$ is not simply absorbing the observed changes in $\Gamma$ and $\kappa$. We also observe that $\nu$ is largely independent of density.

Number Loss over Measurement

As we measure the contrast as a function of time, we also have access to the number loss over the measurement, which can be significant. We first fit the number decay in the deep lattice at each electric field to an exponential decay, $N_{\text{mol}}(T) = N_0 e^{-\alpha T}$, as the loss is dominated by off-resonant light scattering (*60*) in the pinned case. For other lattice conditions, we fit the number decay to a two-body plus one-body model $N_{\text{mol}}(T) = N_0\alpha'/(-N_0\beta + e^{\alpha' T}(\alpha' + N_0\beta))$, where the one-body rate $\alpha'$ is $\alpha$ scaled by the relative ratio of lattice depths to the deep lattice case, that is $\alpha' = \alpha(V_{\text{H1}} + V_{\text{H2}} + V_{\text{V}})/(3 \times 65)$ where $(V_{\text{H1}}, V_{\text{H2}}, V_{\text{V}})$ is the lattice depth in recoils in the first horizontal direction, the second horizontal direction, and the vertical direction respectively. We use the extracted $N_0$ from the fit to calculate the initial density as $n_0 = N_0/(19 \times 4\pi\sigma_{x'}\sigma_{z'})$, where $\sigma_{x'}, \sigma_{z'}$ are the starting cloud sizes. We can also examine the fitted two-body loss rate $\beta$ as a function of lattice depth, shown as Fig. S2. Loss is lowest when there is little molecular dephasing around $\chi = 0$ Hz, but it is similar between $\chi = -205$ Hz and $\chi = 102$ Hz where it only decreases with increasing lattice depth. We also note that loss is slightly lower in 1D than 2D and attribute the difference to a reduction of collisional partners in each tube.

Dynamical Decoupling and Rabi Frequencies

For our dynamical decoupling, we used a symmetric KDD sequence (*40*) which consists of 10 pulses spaced by 50 $\mu$s to make a block of total length 0.5 ms. We used rectangular pulses with Rabi frequencies of approximately 100 kHz. We chose to use KDD pulse sequences rather than XY8 pulse sequences used in (*22*) to be less sensitive to pulse error, and we found that KDD with the 20 kHz filter frequency works well to remove single particle dephasing, which can be time-dependent arising from fluctuations in dc electric fields and motion in the optical trap.

Extended MACE (EMACE)

To quantitatively model the dynamics of itinerant dipoles, we develop an extended moving-average cluster expansion (EMACE) method, which accounts for the motion of the molecules, including associated collision and loss processes. Detailed description of the EMACE is provided in the Supplementary Text. In brief, for each molecule initially localized at a site $i$, we solve for the itinerant dynamics in a "buffer zone" consisting of lattice sites connected to $i$, via the tunneling matrix, which may include other initial molecules as well. In addition, for every site in the buffer zone, we also include the $M$ most strongly coupled neighboring molecules. This technique enables us to directly compute the molecular dynamics as the lattice depth is varied. As shown in Fig. 4, this method enables us to observe the experimental trends in the contrasts decay rate as the lattice depth is lowered in both 1D and 2D geometries as the molecules begin to move in the lattice.

Scattering Theory for Molecules in a 2D Layer

Here, we provide a condensed version of the information presented in the Supplementary Text regarding the scattering theory presented in this work. In the fully itinerant regime, a nondegenerate gas of KRb molecules has its mean kinetic energy per particle much larger than the long-range dipole-dipole interaction energy. Decay of the Ramsey contrast is, therefore, primarily caused by molecule-molecule collisions that decohere the one-body molecular states. The result of each collision can be stated concisely: a collision causes each scattering channel to accrue a different scattering phase shift, leading to decoherence of the post-collision reduced density matrix of each molecule upon tracing over the other. For two molecules $A$ and $B$ that have collided, a dilute gas makes it highly unlikely that any third molecule then colliding with $B$, or its subsequent collision partners, would later collide again with $A$. This argument is made stronger by the fact that subsequent collisions of molecules increase the likelihood of scattering in the singlet channel, which is known to lead to large $s$-wave losses (*18*). Our theoretical model proceeds with this Markov approximation, which warrants us keeping track of only the single molecule reduced density matrices since subsequent collisions with molecule $B$, previously entangled with molecule $A$, cannot reduce the purity of molecule $A$ (supplementary material of (*61*)). The interplay of motion and coherent collisions is well captured by Monte Carlo simulations of classically itinerant molecules in 2D, subject to a collisional decoherence model described below.

First initializing a thermal ensemble of molecules in a 2D layer, our simulation evolves their harmonically confined trajectories in classical phase space with a standard Störmer-Verlet integration scheme (*62*). Pairwise collisions are then sampled from a probability distribution based on the joint two-molecule spin states and their relative momenta. If determined to collide inelastically, the molecules are assumed to be lost from the trap and discarded from the simulation. Conversely, elastic collisions update the individual molecular states by 1) imparting channel-dependent scattering phase shifts on the joint two-body molecular state, and 2) tracing over a molecule to give the resulting one-body reduced density matrices. Such a procedure amounts to a noisy quantum channel on each of the scattered molecules, leading to a loss of coherence and reduction of the Ramsey contrast. The Ramsey contrast can be extracted from the simulation ensemble at any given simulation time step. See the Supplementary Text for further details of our simulation. The extracted contrast decay rates from our simulations compare favorably with the experimental data as shown in Fig. 3B, assuring us that our theory does indeed incorporate the relevant physics.

For a more intuitive understanding of the observed collisional decoherence, we derive an analytic model that estimates the contrast decay rate as a function of the electric field. For two molecules prepared in an equal superposition of their internal rotational states $|\uparrow\rangle$ and $|\downarrow\rangle$ the change in their contrast following a single collision is derived to be $\Delta C = 2\sin^2(\delta_{\leftrightarrow} - \delta_{\updownarrow})$. Then assuming the gas is dilute enough such that each molecule will suffer, at most, only a single collision, we approximate the contrast decay rate $\Gamma$ as the linear slope over which the contrast changes at the elastic collision rate $\beta_{\mathrm{el}} = (\beta_{\leftrightarrow} + \beta_{\updownarrow})/2$: $\Gamma = \beta_{\mathrm{el}}\Delta C$. These quantities are then integrated against a Maxwell-Boltzmann distribution to give a thermally averaged contrast decay rate $\langle\Gamma\rangle$, and $\kappa \approx \langle\Gamma\rangle/\, n$. This model allows a treatment of both quasi-2D and 1D geometries, but its simplifications grant it only qualitative power in describing the trends of $\kappa$ with electric field and $\chi$. The Supplementary Text provides explicit derivations of the estimated $\langle\Gamma\rangle$, in both one and two dimensions.

Two-Body Theory Model in a Lattice:

For shallow lattices, we also model the contrast decay by combining the dynamics of pairs of molecules. For a given molecule $A$, we first independently compute the two-body dynamics of this molecule with each other molecule $B$ to get a two-body estimate for the contrast $\langle 2\hat{s}_A^{X(A,B)}\rangle(T)$, where we move into a rotating frame such that $\langle \hat{s}_A^{Y}(T)\ \rangle\ =\ 0$. We then compute the combined contrast by multiplying the contrast values for each pair $\langle 2\hat{s}_A^{X}\rangle(T), = \prod_B \langle 2\hat{s}_A^{X(A,B)}\rangle(T)$. This is inspired by a scattering picture, where each collision independently reduces the contrast by a given fraction and can be more formally related to a trotterization of the Hamiltonian (see Supplementary Text for details). Consequently, this approach is valid for shallow lattices, where molecules are sufficiently fast such that the dynamics of each pair can be treated independently, while the other pairs only lead to background decoherence, akin to a Markovian bath. The results of this simulation are shown for tunneling rates greater than 50 Hz in Fig. 4A and Fig. 4C.

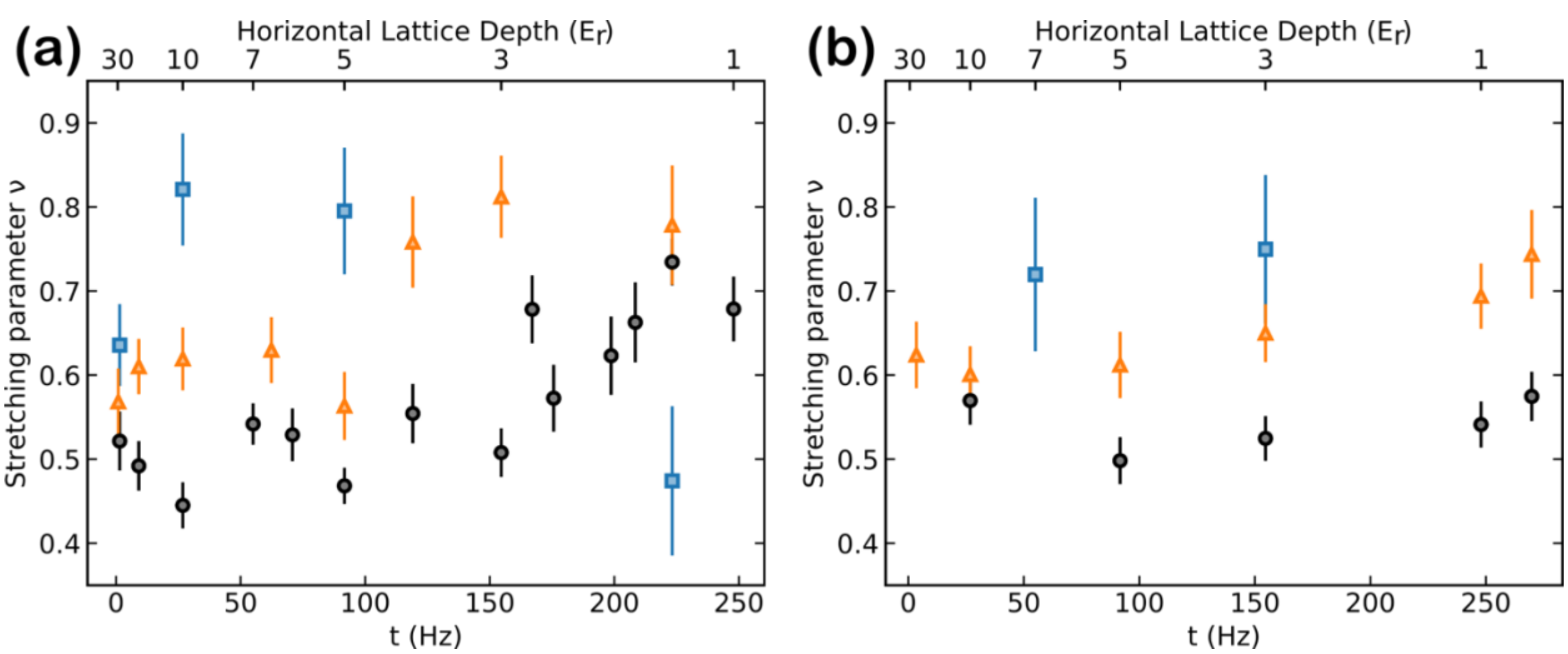

**Fig. S1: Extracted mean stretching parameter for itinerant molecules**. **a)** Fitted stretching parameter $\nu$ averaged over all densities at each molecular tunneling rate $t$ in two directions. Black

circles, blue squares, and orange triangles present experimental data for $\chi = -205, 0,$ 102 Hz respectively. Error bars are averaged errors of 1 s.e. to stretched exponential fit. **b)** Same as a) but for molecular tunneling only in one direction.

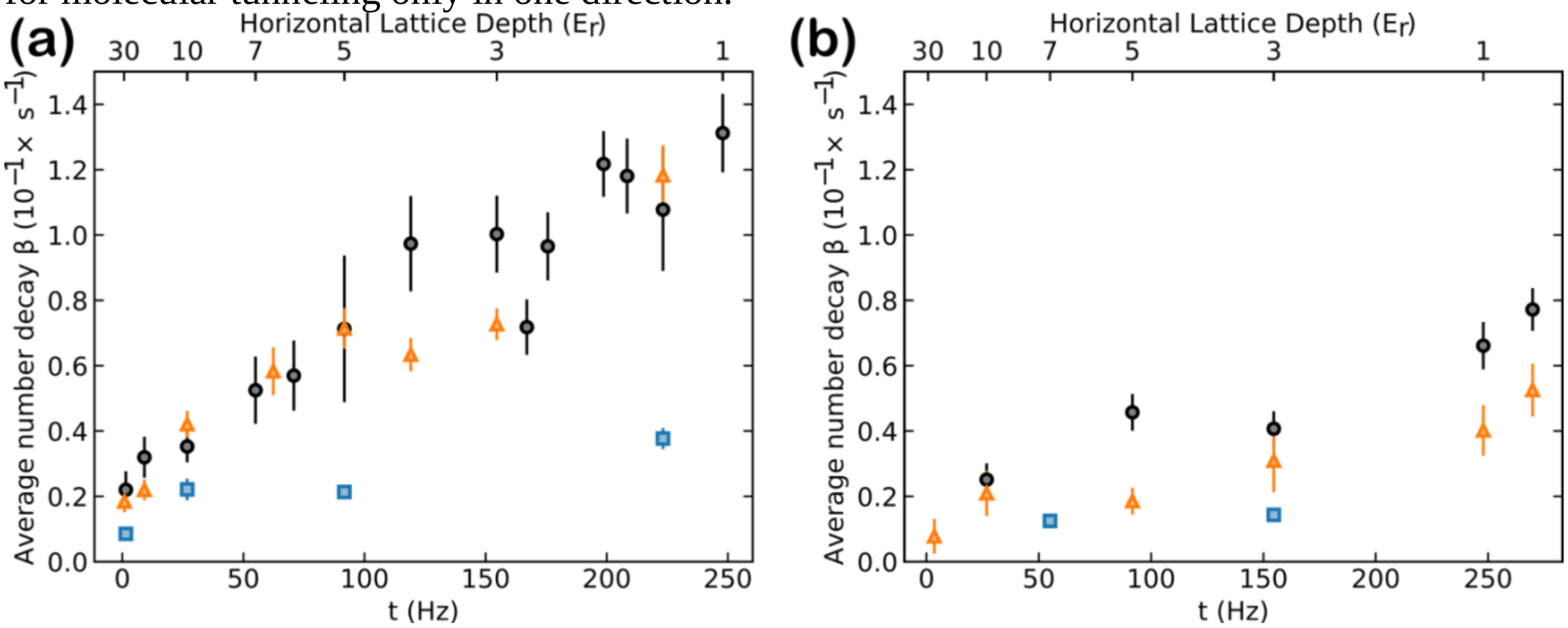

**Fig. S2: Extracted mean two-body loss rate for itinerant molecules**. **a)** Fitted two-body loss rate $\beta$ averaged over all densities at each molecular tunneling rate $t$ in two directions. Black circles, blue squares, and orange triangles present experimental data for $\chi = -205, 0,$ 102 Hz respectively. Error bars are averaged errors of 1 s.e. to two-body plus one-body loss model **b)** Same as a) but for molecular tunneling only in one direction.

# Supplementary Text

## CONTENTS

## I. COLLISIONAL DECOHERENCE

In the main text, we describe our observation of coherent spin dynamics across a wide range of allowed molecular motion: from being completely restricted by a deep optical lattice, to being fully itinerant. In this supplemental document, we present details of our various theoretical models, each of which captures the dominant physics of these different regimes, along with additional supporting information relevant to the experiment.

We commence discussions with the case of fully itinerant KRb molecules (i.e. no applied lattice potential, $V_{L,\alpha} = 0$ for all $\alpha$). In this regime, collisional decoherence of KRb molecules is modeled with a microscopic theory of scattering on the dipolar interaction Hamiltonian from Eq. (1) of the main text, but handled in first quantization. For an appropriately applied magnetic field and a sufficiently large electric field, we treat the molecules as rigid rotors (8) where the dipoles are always polarized along the field axis with $m_N = 0$. Any explicit references to $m_N$ will thus be dropped in what follows. The dominant dipole-dipole interaction potential between molecules $A$ and $B$ is, therefore, well approximated as

$$V_{\rm dd}(\boldsymbol{r}) = \langle N_A, N_B|\widehat{V}_{\rm dd}(\boldsymbol{r})|N_A', N_B'\rangle = \langle N_A|\widehat{d}_A|N_A'\rangle\langle N_B|\widehat{d}_B|N_B'\rangle \frac{1 - 3(\hat{\boldsymbol{r}}\cdot\hat{\boldsymbol{E}})^2}{4\pi\epsilon_0 r^3}, \tag{S1}$$

with the dipole matrix elements corresponding to the field-dressed (labeled without tildes) states

$$\langle N|d|N'\rangle = d\sum_{\tilde{N},\tilde{N}'}\langle N,0|\tilde{N},0\rangle\langle\tilde{N}',0|N',0\rangle\sqrt{(2\tilde{N}+1)(2\tilde{N}'+1)}\begin{pmatrix}\tilde{N} & 1 & \tilde{N}'\\ 0 & 0 & 0\end{pmatrix}^2. \tag{S2}$$

In the experiment, only $|{\downarrow}\rangle = |N=0\rangle$ and $|{\uparrow}\rangle = |N=1\rangle$ are addressed, which we take to be the relevant states for scattering as well. Utilizing only these states then identifies 4 relevant asymptotic scattering channels:

$$|{\Downarrow}\rangle = |{\downarrow},{\downarrow}\rangle\,, \tag{S3a}$$

$$|{\leftrightarrow}\rangle = \frac{|{\downarrow},{\uparrow}\rangle + |{\uparrow},{\downarrow}\rangle}{\sqrt{2}}, \tag{S3b}$$

$$|{\Uparrow}\rangle = |{\uparrow},{\uparrow}\rangle\,, \tag{S3c}$$

$$|{\circ}\rangle = \frac{1}{\sqrt{2}}\left(|{\downarrow},{\uparrow}\rangle - |{\uparrow},{\downarrow}\rangle\right), \tag{S3d}$$

where the first 3 states above reside in the symmetric spin sector, while the last is antisymmetric under particle exchange. Although initially individually prepared as a coherent superposition of $|{\downarrow}\rangle$ and $|{\uparrow}\rangle$, interactions over time could decohere the reduced density matrix of any one molecule. Therefore, the state of a single molecule at any given time is most generally written in terms of a density matrix:

$$\varrho = \sum_{\mu,\nu} |\mu\rangle\, \varrho_{\mu,\nu}\, \langle\nu|\,, \tag{S4}$$

where Greek indices label $\nu = \Downarrow, \Uparrow, \leftrightarrow, \circ$, and $\sum_\nu \varrho_{\nu,\nu} = 1$. Following a collision occurrence, each of the scattering channels accrues a scattering phase shift that is imparted via the $S$-matrix, $\hat{S}$. The 2-molecule density matrix following a collision is then written as

$$\varrho' = \hat{S}\varrho\hat{S}^\dagger. \tag{S5}$$

In our model, we treat all inelastic processes as leading to molecular loss, leaving a strictly diagonal $S$-matrix

$$S_{\nu,\nu} = e^{2i\delta_\nu(k)}, \tag{S6}$$

but with complex-valued channel-dependent phase shifts $\delta_\nu(k)$.

The $S$-matrix above then treats each channel as effectively separate during a collision, with the only relevant dipole matrix elements being the diagonal ones:

$$\langle{\Downarrow}|\, \widehat{V}_{\rm dd}(\boldsymbol{r})\, |{\Downarrow}\rangle = \frac{d_\downarrow^2}{4\pi\epsilon_0}\left(\frac{1-3(\hat{\boldsymbol{r}}\cdot\hat{\boldsymbol{E}})^2}{r^3}\right), \tag{S7a}$$

$$\langle{\leftrightarrow}|\, \widehat{V}_{\rm dd}(\boldsymbol{r})\, |{\leftrightarrow}\rangle = \frac{(d_\downarrow d_\uparrow + d_{\downarrow\uparrow}^2)}{4\pi\epsilon_0}\left(\frac{1-3(\hat{\boldsymbol{r}}\cdot\hat{\boldsymbol{E}})^2}{r^3}\right), \tag{S7b}$$

$$\langle{\Uparrow}|\, \widehat{V}_{\rm dd}(\boldsymbol{r})\, |{\Uparrow}\rangle = \frac{d_\uparrow^2}{4\pi\epsilon_0}\left(\frac{1-3(\hat{\boldsymbol{r}}\cdot\hat{\boldsymbol{E}})^2}{r^3}\right) \tag{S7c}$$

$$\langle{\circ}|\, \widehat{V}_{\rm dd}(\boldsymbol{r})\, |{\circ}\rangle = \frac{(d_\downarrow d_\uparrow - d_{\downarrow\uparrow}^2)}{4\pi\epsilon_0}\left(\frac{1-3(\hat{\boldsymbol{r}}\cdot\hat{\boldsymbol{E}})^2}{r^3}\right), \tag{S7d}$$

having identified the effective dipole moments with the notation

$$d_\downarrow \equiv \langle{\downarrow}|\, \widehat{d}_y\, |{\downarrow}\rangle = \langle N=0|\, \widehat{d}_y\, |N=0\rangle\,, \tag{S8a}$$

$$d_\uparrow \equiv \langle{\uparrow}|\, \widehat{d}_y\, |{\uparrow}\rangle = \langle N=1|\, \widehat{d}_y\, |N=1\rangle\,, \tag{S8b}$$

$$d_{\downarrow\uparrow} \equiv \langle{\downarrow}|\, \widehat{d}_y\, |{\uparrow}\rangle = \langle{\uparrow}|\, \widehat{d}_y\, |{\downarrow}\rangle = \langle N=1|\, \widehat{d}_y\, |N=0\rangle\,. \tag{S8c}$$

We also define the dipole lengths (48) for each scattering channel as

$$a_D^{\Downarrow} = \frac{\mu d_{\Downarrow}^2}{4\pi\epsilon_0\hbar^2} = \frac{\mu d_\downarrow^2}{4\pi\epsilon_0\hbar^2}, \tag{S9a}$$

$$a_D^{\leftrightarrow} = \frac{\mu d_+^2}{4\pi\epsilon_0\hbar^2} = \frac{\mu(d_\downarrow d_\uparrow + d_{\downarrow\uparrow}^2)}{4\pi\epsilon_0\hbar^2}, \tag{S9b}$$

$$a_D^{\Uparrow} = \frac{\mu d_{\Uparrow}^2}{4\pi\epsilon_0\hbar^2} = \frac{\mu d_\uparrow^2}{4\pi\epsilon_0\hbar^2}, \tag{S9c}$$

$$a_D^{\circ} = \frac{\mu d_-^2}{4\pi\epsilon_0\hbar^2} = \frac{\mu(d_\downarrow d_\uparrow - d_{\downarrow\uparrow}^2)}{4\pi\epsilon_0\hbar^2}, \tag{S9d}$$

where $\mu$ is the reduced mass. With low energy collisions off a repulsive potential, most of the collision occurs in the long-range $1/r^3$ interaction tail. As such, the collision dynamics is taken to occur on time scales slow compared to the dynamical decoupling pulses, rendering the effective dipole lengths in channels $|\Downarrow\rangle$ and $|\Uparrow\rangle$ to be time averaged such that

$$a_D^{\updownarrow} \equiv \langle a_D^{\Downarrow} \rangle_t = \langle a_D^{\Uparrow} \rangle_t = \frac{\mu(d_{\downarrow}^2 + d_{\uparrow}^2)}{8\pi\epsilon_0\hbar^2}, \tag{S10}$$

where $\langle \ldots \rangle_t$ denotes a time averaging over the decoupling pulses. We utilize the dipole length from aligned (direct) interactions $a_D^{\updownarrow}$, in place of $a_D^{\Downarrow}$ and $a_D^{\Uparrow}$ in the remainder of what follows.

To justify our assumption that 2-body collisions are truly the dominant form of dipolar physics, we compare the thermal energy $k_B T_{\text{avg}}$ to the average dipolar mean-field energy per particle, computed analytically for a trapped thermal gas of dipoles as

$$\varepsilon_{\text{mf}} = -\frac{N_{\text{mol}}}{6\sqrt{2\pi}} \frac{d_{\text{max}}^2}{4\pi\epsilon_0 \sigma_\perp^2 a_{\text{ho}}} F\left( \frac{\sqrt{2}\sigma_\perp}{a_{\text{ho}}} \right), \tag{S11a}$$

where $a_{\text{ho}} = \sqrt{\hbar/(m\omega_{\text{ho},y})}$ is the single-molecule harmonic oscillator length along the tightly confined $y$-direction, $\sigma_\perp = \sqrt{k_B T_{\text{avg}}/(m\omega_{\text{ho},\perp}^2)}$ is the thermal length in the transverse directions and $F(x)$ is the molecular cloud anisotropy function (63). At temperatures of $T_{\text{avg}} = 300$ nK and molecule numbers of $N_{\text{mol}} = 500$ seen in the experiment, we find that $\varepsilon_{\text{mf}}/(k_B T_{\text{avg}}) \lesssim 0.01$, allowing us to safely ignore beyond two-body scattering processes. We ensure this approximation is valid in the explored parameter regime by using a conservative estimate where $\varepsilon_{\text{mf}}$ is computed with the largest induced dipole moment in KRb at any given electric field, $d_{\text{max}}^2 = \max\{d_{\Downarrow}^2, d_{\Uparrow}^2, d_{\leftrightarrow}^2, d_{\circ}^2\}$.

### A. Itinerant collision dynamics in quasi-2D

For collisions that occur in quasi-2D at energies close-to-threshold, the scattering cross sections from a dipolar potential are well approximated by those obtained in Born approximation (48). Under this approximation, the diagonal $S$-matrix elements are given as (64)

$$S_{\nu,\nu}^{\text{2D}}(\boldsymbol{k}', \boldsymbol{k}) = e^{2i\delta_\nu^{\text{2D}}(\boldsymbol{k}',\boldsymbol{k})} = 1 - 2\pi i \left\langle \boldsymbol{k}' \right| \int dy |\varphi_0(y)|^2 \langle \nu | \widehat{V}_{\text{dd}}(\boldsymbol{r}) | \nu \rangle \left| \boldsymbol{k} \right\rangle \rho_E(E_k), \tag{S12}$$

with $|\boldsymbol{k}'| = |\boldsymbol{k}|$ and assuming the wavefunction along $y$ remains in its harmonic oscillator ground state $\varphi_0(y)$. Above, $\rho_E(E_k)$ is the density of states with energy $E_k$, which for our treatment of collisions in free-space here, is absorbed into the antisymmetrized ($\mathcal{A}$) and symmetrized ($\mathcal{S}$) momentum eigenstates:

$$|\boldsymbol{k}\rangle_{\mathcal{A}} = i\frac{\sqrt{\mu}}{\pi\hbar}\frac{\sin(\boldsymbol{k}\cdot\boldsymbol{\rho})}{\sqrt{2}}, \tag{S13a}$$

$$|\boldsymbol{k}\rangle_{\mathcal{S}} = \frac{\sqrt{\mu}}{\pi\hbar}\frac{\cos(\boldsymbol{k}\cdot\boldsymbol{\rho})}{\sqrt{2}}, \tag{S13b}$$

appropriate to the triplet and singlet scattering channels respectively. That is to say, the states defined above satisfy the energy-normalization condition

$${}_{\mathcal{A}}\left\langle \boldsymbol{k} \middle| \boldsymbol{k}' \right\rangle_{\mathcal{A}} = {}_{\mathcal{S}}\left\langle \boldsymbol{k} \middle| \boldsymbol{k}' \right\rangle_{\mathcal{S}} = \delta(E_k - E_k')\delta(\phi_k - \phi_k'), \tag{S14}$$

where $\phi_k$ specifies the incident relative momentum direction. We use the notation of $\boldsymbol{\rho} = (z', x')$ and $\boldsymbol{k} = (k_{z'}, k_{x'})$ as the in-plane position and momentum coordinates (refer to the main text for the lab-frame axes). Evaluating the necessary integrals, we obtain the matrix elements analytically as

$$\begin{aligned} {}_{\mathcal{A}}\left\langle \nu; \boldsymbol{k}' \right| \int dy |\varphi_0(y)|^2 \widehat{V}_{\text{dd}}(\boldsymbol{r}) \left| \nu; \boldsymbol{k} \right\rangle_{\mathcal{A}} = \frac{k a_D^\nu}{\pi} e^{k^2 a_{\text{ho}}^2 \xi_-(\phi_s)} \Big[ &\sqrt{\xi_+(\phi_s)} e^{k^2 a_{\text{ho}}^2 \cos\phi_s} \text{Erfc}\left( k a_{\text{ho}} \sqrt{\xi_+(\phi_s)} \right) \\ &- \sqrt{\xi_-(\phi_s)} \text{Erfc}\left( k a_{\text{ho}} \sqrt{\xi_-(\phi_s)} \right) \Big], \end{aligned} \tag{S15a}$$

$$\begin{aligned} {}_{\mathcal{S}}\left\langle \nu; \boldsymbol{k}' \right| \int dy |\varphi_0(y)|^2 \widehat{V}_{\text{dd}}(\boldsymbol{r}) \left| \nu; \boldsymbol{k} \right\rangle_{\mathcal{S}} = \frac{k a_D^\nu}{\pi} \Bigg[ \frac{4}{3\sqrt{\pi}} \frac{1}{k a_{\text{ho}}} - e^{k^2 a_{\text{ho}}^2 \xi_-(\phi_s)} \Bigg( &\sqrt{\xi_+(\phi_s)} e^{k^2 a_{\text{ho}}^2 \cos\phi_s} \text{Erfc}\left( k a_{\text{ho}} \sqrt{\xi_+(\phi_s)} \right) \\ &+ \sqrt{\xi_-(\phi_s)} \text{Erfc}\left( a_{\text{ho}} k \sqrt{\xi_-(\phi_s)} \right) \Bigg) \Bigg], \end{aligned} \tag{S15b}$$

where $\xi_{\pm}(\phi_s) = (1 \pm \cos\phi_s)/2$ is a function of the scattering angle $\phi_s = \cos^{-1} \hat{\boldsymbol{k}} \cdot \hat{\boldsymbol{k}}'$ and $\mathrm{Erfc}(z)$ is the complementary error function. The phase shifts are, therefore, identified as:

$$e^{2i\delta_\nu^{\mathrm{2D}}(k,\phi_s)} \approx 1 - 2i(ka_D^\nu)e^{k^2 a_{\mathrm{ho}}^2 \xi_-(\phi_s)} \left[\sqrt{\xi_+(\phi_s)}e^{k^2 a_{\mathrm{ho}}^2 \cos\phi_s}\mathrm{Erfc}\left(ka_{\mathrm{ho}}\sqrt{\xi_+(\phi_s)}\right) - \sqrt{\xi_-(\phi_s)}\mathrm{Erfc}\left(ka_{\mathrm{ho}}\sqrt{\xi_-(\phi_s)}\right)\right], \quad \nu = \Downarrow, \Uparrow, \leftrightarrow, \tag{S16a}$$

$$e^{2i\delta_\circ(k,\phi_s)} \approx 1 - 2i(ka_D^\circ)\left[\frac{4}{3\sqrt{\pi}}\frac{1}{ka_{\mathrm{ho}}} - e^{k^2 a_{\mathrm{ho}}^2 \xi_-(\phi_s)}\left(\sqrt{\xi_+(\phi_s)}e^{k^2 a_{\mathrm{ho}}^2 \cos\phi_s}\mathrm{Erfc}\left(ka_{\mathrm{ho}}\sqrt{\xi_+(\phi_s)}\right) + \sqrt{\xi_-(\phi_s)}\mathrm{Erfc}\left(a_{\mathrm{ho}}k\sqrt{\xi_-(\phi_s)}\right)\right)\right]. \tag{S16b}$$

From the expressions above, we see that the low energy 2D elastic cross section scales as $\sigma_{\mathrm{2D}} \sim k(a_D^\nu)^2$, consistent with Ref. (48). With a strong preference for forward scattering at the temperatures considered here (also observed in Ref. (65) with full scattering calculations), the elastic and inelastic scattering rates are well approximated by

$$\beta_\nu^{\mathrm{2D,el}}(k) \approx n_{\mathrm{2D}}\frac{4\pi\hbar}{\mu}\left|1 - e^{2i\delta_\nu^{\mathrm{2D}}(k,0)}\right|^2, \tag{S17a}$$

$$\beta_\nu^{\mathrm{2D,inel}}(k) \approx n_{\mathrm{2D}}\frac{4\pi\hbar}{\mu}\left(1 - \left|e^{2i\delta_\nu^{\mathrm{2D}}(k,0)}\right|^2\right), \tag{S17b}$$

giving the total scattering rate $\beta_\nu^{\mathrm{2D}}(k) = \beta_\nu^{\mathrm{2D,el}}(k) + \beta_\nu^{\mathrm{2D,inel}}(k)$ (37), where $n_{\mathrm{2D}}$ is the average 2D planar density. The expressions above are calculated for identical fermions with the dipoles oriented orthogonal to the plane of free motion.

### 1. *Monte Carlo simulations of itinerant collisional KRb*

Equipped with these scattering phase shifts, time traces of the contrast in a fully itinerant sample can now be generated with numerical Monte Carlo simulations. To do so, we initialize a Maxwell-Boltzmann distributed ensemble of harmonically confined molecules in 2D, all identically prepared in state $|\psi_0\rangle = (|\downarrow\rangle + |\uparrow\rangle)/\sqrt{2}$. The molecules are taken to undergo motion in classical phase space, progressing forward in discrete time steps of $\Delta t$ via Störmer-Verlet symplectic integration (62) in the presence of the background optical dipole trap (ODT). The ODT is well approximated by a harmonic potential

$$V_{\mathrm{trap}}(\boldsymbol{\rho}) = \frac{1}{2}m(\omega_{\mathrm{ho},x'}^2 x'^2 + \omega_{\mathrm{ho},z'}^2 z'^2) \tag{S18}$$

with harmonic trapping frequencies $\omega_{\mathrm{ho},\alpha'}$, which will also be relevant to discussions of lattice dynamics later on. Primes on coordinate labels are to distinguish between the axes set by the ODT, and those imposed by the applied lattice (see Methods). Alongside the molecular positions $\boldsymbol{\rho}_k$ and momenta $\boldsymbol{p}_k$, we also keep track of the reduced density matrix $\boldsymbol{\varrho}$ of each molecule, ignoring the quantum correlations between molecules. We find this tracking of only single-molecule states a valid assumption since subsequent collisions of a molecule $B$, previously collisionally entangled with another molecule $A$, cannot decrease the reduced density matrix purity of $A$ (supplementary material of Ref. (61). This statement is true so long as the subsequent collision partners of $B$, or $B$ itself, do not re-collide with $A$, which is a good approximation in dilute gases.

Collisions are sampled using the direct simulation Monte Carlo (DSMC) method (66-68), which exploits the locality of interactions for computational efficiency. In our implementation, the simulation volume is first partitioned into discrete grid cells of volume $\Delta V_{\mathrm{cell}}$, into which the simulated molecules are binned based on their positions. Collisions are then assumed to only occur within each grid cell with probability

$$P_{\mathrm{coll}}(k) = \frac{\Delta t}{\Delta V_{\mathrm{cell}}}\sum_\nu \rho_{\nu,\nu}^{\mathrm{2D}}\beta_\nu(k), \tag{S19}$$

which depends on the relative momentum $|\boldsymbol{p}_r| = |\boldsymbol{p}_A - \boldsymbol{p}_B| = \hbar k$, and the appropriately symmetrized 2-body density matrix $\rho = \{\varrho_A \otimes \varrho_B\}_{\mathrm{sym}}$ of molecules $A$ and $B$. Curly braces $\{\ldots\}_{\mathrm{sym}}$ denote a transformation from the basis $\{|\downarrow\downarrow\rangle, |\downarrow\uparrow\rangle, |\uparrow\downarrow\rangle, |\uparrow\uparrow\rangle\}$, into the basis of Eq. (S3). If determined to occur, the collision must be assigned as elastic or inelastic, which is done as follows.

Given that all $p$-wave ($|m_{\rm pw}| = 1$, where $m_{\rm pw}$ denotes a 2D partial wave) losses are far suppressed over $s$-wave ($m_{\rm pw} = 0$) ones (18), we will treat all scattering phase shifts in the symmetric sector as real-valued (i.e. we only take their real parts if complex). As for the antisymmetric channel, scattering of identical fermions in $|\circ\rangle$ must necessarily involve the $m_{\rm pw} = 0$ partial wave, which could result in short-range inelastic loss. As such, we model scattering in the antisymmetric sector with a complex phase shift $\delta_{\rm anti} = \delta_\circ + i\eta_s$, comprising a real-valued dipolar part $\delta_\circ$ (S16b), and an imaginary $m_{\rm pw} = 0$ contribution $i\eta_s$. The purely imaginary $s$-wave phase shift is motivated by the observed universal short-range loss in KRb (69,70). By careful comparison with the experimentally observed number loss rates, we simply insert an empirically determined imaginary phase shift of $\eta_s = 0.05$ that reproduces it. This value implies about a 20% probability of loss for collisions in the singlet channel. The resultant inelastic scattering rate is then computed as

$$\beta_\circ^{\rm 2D,inel} \approx n_{\rm 2D} \frac{4\pi\hbar}{\mu} \left(1 - e^{-4\eta_s}\right), \tag{S20}$$

so that the total scattering rate in channel $|\circ\rangle$ is given as $\beta_\circ^{\rm 2D} = \beta_\circ^{\rm 2D,el} + \beta_\circ^{\rm 2D,inel}$. The probability that a simulated collision occurrence is inelastic is then taken to be

$$P_{\rm inel} = \frac{\varrho_{\circ,\circ}\beta_\circ^{\rm 2D,inel}}{\varrho_{\Downarrow,\Downarrow}\beta_\Downarrow^{\rm 2D} + \varrho_{\Uparrow,\Uparrow}\beta_\Uparrow^{\rm 2D} + \varrho_{\leftrightarrow,\leftrightarrow}\beta_\leftrightarrow^{\rm 2D} + \varrho_{\circ,\circ}\beta_\circ^{\rm 2D}}, \tag{S21}$$

and treated as resulting in trap-loss of the molecular pair. Trap-loss translates to a discarding of these molecules in the simulation.

If elastic, however, the collision must modify the reduced density matrix. First, a strong preference for forward scattering (S16) allows us to approximate differential scattering by sampling the scattering angle as $\phi_s = 0$ or $\pi$ with equal probability, to produce $\boldsymbol{p}_r' = +\boldsymbol{p}_r$ or $-\boldsymbol{p}_r$ respectively. We leave consideration of the full angle dependence of $\delta_\nu^{\rm 2D}$ to a future more detailed study. Then taking the symmetrized 2-body density matrix $\varrho$, we apply only the elastic scattering phase shifts to it:

$$\varrho' = \sum_{\mu,\nu} e^{2i\delta_\mu^{\rm 2D}} |\mu\rangle \, \varrho_{\mu,\nu} \, \langle\nu| \, e^{-2i\delta_\nu^{\rm 2D}}. \tag{S22}$$

Then adopting the basis ordering in Eq. (S3), we perform a partial trace over the second molecule $B$ which comprises a sum over projectors onto the single-molecule reduced Hilbert space

$${\rm tr}_B\{\ldots\} = \Pi_{\downarrow_B}(\ldots)\Pi_{\downarrow_B}^\dagger + \Pi_{\uparrow_B}(\ldots)\Pi_{\uparrow_B}^\dagger, \tag{S23}$$

where each projector has the matrix representation

$$\Pi_{\downarrow_B} = \begin{array}{c} \begin{array}{cccc} \langle\downarrow_B|\Downarrow\rangle & \langle\downarrow_B|+\rangle & \langle\downarrow_B|\Uparrow\rangle & \langle\downarrow_B|-\rangle \end{array} \\ \begin{pmatrix} 1 & 0 & 0 & 0 \\ 0 & 1/\sqrt{2} & 0 & -1/\sqrt{2} \end{pmatrix} \begin{array}{c} |\downarrow_A\rangle \\ |\uparrow_A\rangle \end{array} \end{array}, \tag{S24a}$$

$$\Pi_{\uparrow_B} = \begin{array}{c} \begin{array}{cccc} \langle\uparrow_B|\Downarrow\rangle & \langle\uparrow_B|+\rangle & \langle\uparrow_B|\Uparrow\rangle & \langle\uparrow_B|-\rangle \end{array} \\ \begin{pmatrix} 0 & 1/\sqrt{2} & 0 & 1/\sqrt{2} \\ 0 & 0 & 1 & 0 \end{pmatrix} \begin{array}{c} |\downarrow_A\rangle \\ |\uparrow_A\rangle \end{array} \end{array}. \tag{S24b}$$

The resulting post-collision reduced density matrix of molecule $A$ is therefore

$$\varrho_A' = {\rm tr}_B\{\rho'\} = \Pi_{\downarrow_B}\varrho'\Pi_{\downarrow_B}^\dagger + \Pi_{\uparrow_B}\varrho'\Pi_{\uparrow_B}^\dagger, \tag{S25}$$

that will also be the reduced density of molecule $B$, which completes a simulated collision event. To extract the contrast within our simulation, we compute the expectation of the Pauli matrix $\sigma_X$, with respect to each single-molecule density matrix $\langle\sigma_X\rangle_{\varrho(t)} = {\rm tr}\{\sigma_X\varrho(t)\}$. The ensemble averaged contrast is then obtained by taking the mean value of $\langle\sigma_X\rangle_{\varrho(t)}$ over all simulated molecules at any given time $t$.

### 2. *Loss-induced distillation of single-molecule pure states*

In the collisional regime, we find that 2-body losses in the molecular gas strongly contribute to contrast decay dynamics. Collisions in the singlet scattering channel introduced by collisional decoherence are key to this mechanism, as these collisions have a high probability of being inelastic due to the lack of a $p$-wave barrier. In turn, decohered molecules have a higher tendency to be inelastically lost from the trap, which erases knowledge of their collision-generated entanglement from the overall many-body state. Such erasure thus acts to inherently distill quantum coherence of single molecules in the remaining sample, resulting in a suppression of the contrast decay. However, this suppression only takes effect on time scales long enough to permit molecules to undergo more than one collision, and is therefore not captured by a simple rescaling of the contrast decay rate. We see this dynamical-loss effect manifest in Fig. S3, when comparing the contrast evolution from Monte Carlo simulations in a sample with nonzero singlet loss ($\eta_s = 0.05$), to that with zero singlet loss ($\eta_s = 0$). This representative comparison assumes a gas of $N_{\mathrm{mol}} = 400$ molecules at initial temperature $T_{\mathrm{avg}} = 300$ nK, with an applied electric field $E = 12.7$ kV/cm.

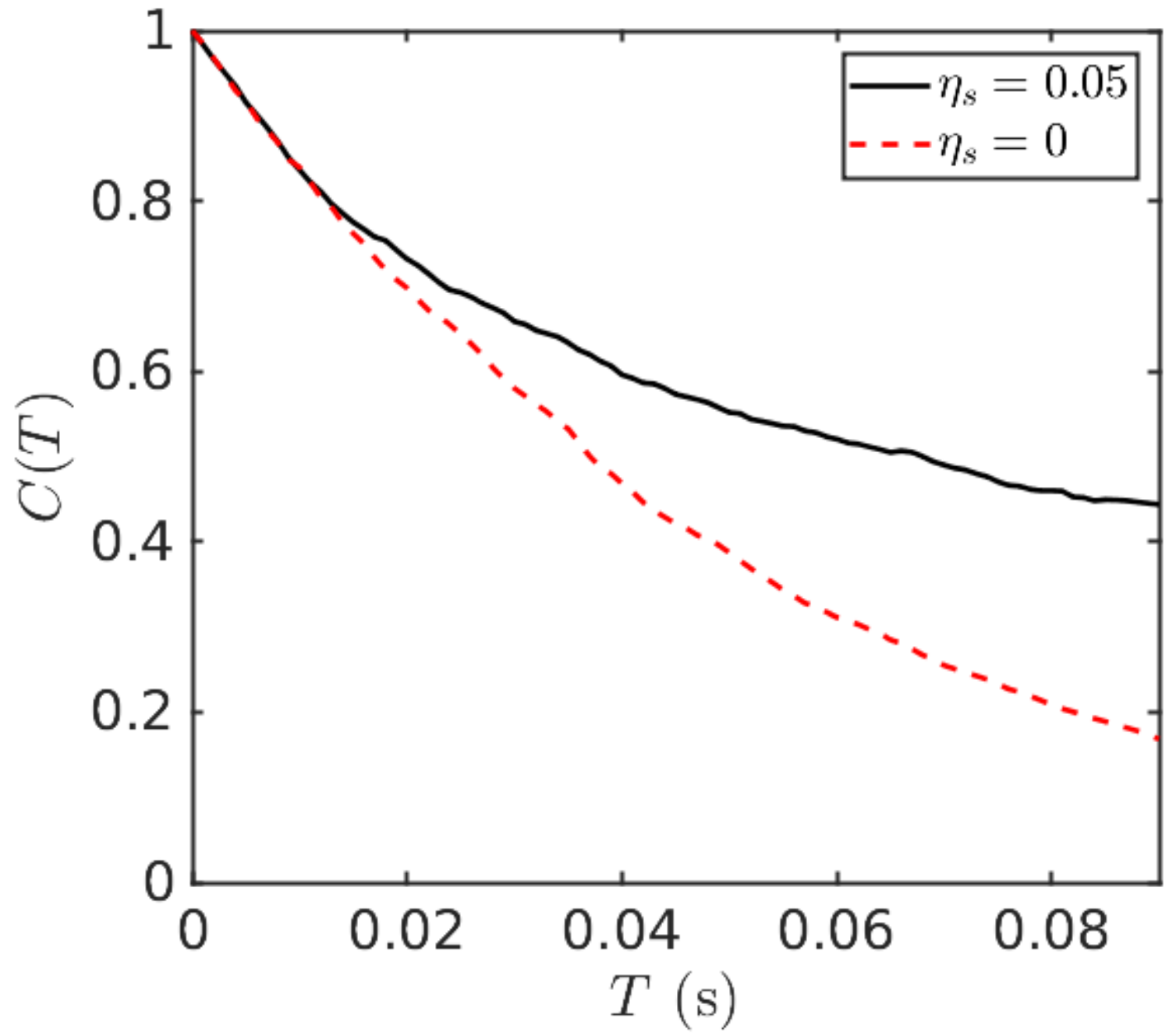

FIG. S3. Contrast decay as a function of time from Monte Carlo simulations including loss (solid black curve with $\eta_s = 0.05$) and no loss (dashed red curve with $\eta_s = 0$). The simulation treats a gas of $N_{\mathrm{mol}} = 400$ molecules at temperature $T_{\mathrm{avg}} = 300$ nK, with applied electric field $E = 12.7$ kV/cm.

In general, the interplay of loss-induced pure state distillation and interaction-induced dephasing could result in a nonlinear density dependence of the contrast decay. Being a two-body process, the presence of loss might even lead to a reduction of contrast decay at larger densities. Although not observed in our current simulations, more complicated density dependences that might arise with more accurate loss models could explain the negative values of $\kappa$ seen in Fig. 3b of the main text. We leave further investigations of this mechanism to future work.

### B. Estimation of the contrast decay

To obtain an estimate of the contrast decay rate, we will consider the decoherence affected by a single collision. Initially prepared in the equal superposition of $|{\downarrow}\rangle$ and $|{\uparrow}\rangle$, the joint quantum state of 2-molecules is given as

$$\begin{aligned} |\psi_{AB}(0)\rangle &= \frac{1}{2}\left(|{\downarrow}\rangle + |{\uparrow}\rangle\right)\left(|{\downarrow}\rangle + |{\uparrow}\rangle\right) \\ &= \frac{1}{2}\left(|{\Downarrow}\rangle + \sqrt{2}\,|{\leftrightarrow}\rangle + |{\Uparrow}\rangle\right). \end{aligned} \tag{S26}$$

Notably, the state above only involves the symmetric spin sector with purely elastic phase shifts, so the treatment here ignores losses. Then following the collision procedure in Sec. I A 1, the post-collision reduced density matrix of the molecules after 1 collision is given by

$$\varrho' = \begin{pmatrix} \frac{1}{2} & \frac{1}{4}e^{-2i(\delta_{\leftrightarrow}+\delta_{\Uparrow})}\left(e^{2i(\delta_{\Downarrow}+\delta_{\Uparrow})}+e^{4i\delta_{\leftrightarrow}}\right) \\ \frac{1}{4}e^{-2i(\delta_{\Downarrow}+\delta_{\leftrightarrow})}\left(e^{2i(\delta_{\Downarrow}+\delta_{\Uparrow})}+e^{4i\delta_{\leftrightarrow}}\right) & \frac{1}{2} \end{pmatrix}, \tag{S27}$$

which gives the contrast in terms of phase shifts as

$$\langle\sigma_X\rangle_{\varrho'} = \cos(\delta_{\Uparrow}-\delta_{\Downarrow})\cos(\delta_{\Uparrow}+\delta_{\Downarrow}-2\delta_{\leftrightarrow}). \tag{S28}$$

See Fig. S4 for a schematic of the different phase shifts picked up in different scattering channels. Noting the dipole length dependence of the phase shifts in Eq. (S16), and the equality of Eq. (S10), the change in contrast after a single collision is therefore

$$\begin{aligned} \Delta C &= 1-\langle\sigma_X\rangle_{\varrho'} \\ &= 2\sin^2\left(\delta_{\updownarrow}-\delta_{\leftrightarrow}\right), \end{aligned} \tag{S29}$$

with $\delta_{\updownarrow} = (\delta_{\Uparrow}+\delta_{\Downarrow})/2$. The contrast decay rate $\Gamma$ is then taken to be the linear slope over which the contrast changes within the time interval $\Delta T = 1/\beta_{\mathrm{el}}$:

$$\Gamma \approx \frac{\Delta C}{\Delta T} = \beta_{\mathrm{el}}\Delta C, \tag{S30}$$

where

$$\beta_{\mathrm{el}} = \frac{1}{4}\left(\beta_{\Downarrow}+2\beta_{\leftrightarrow}+\beta_{\Uparrow}\right). \tag{S31}$$

Since still dependent on $k$, it is more appropriate to consider the thermally averaged contrast decay rate $\langle\Gamma\rangle$, obtained by integrating $\beta_{\mathrm{el}}(k)$ and $\Delta C(k)$ over the equilibrium Maxwell-Boltzmann distribution individually, then taking their product.

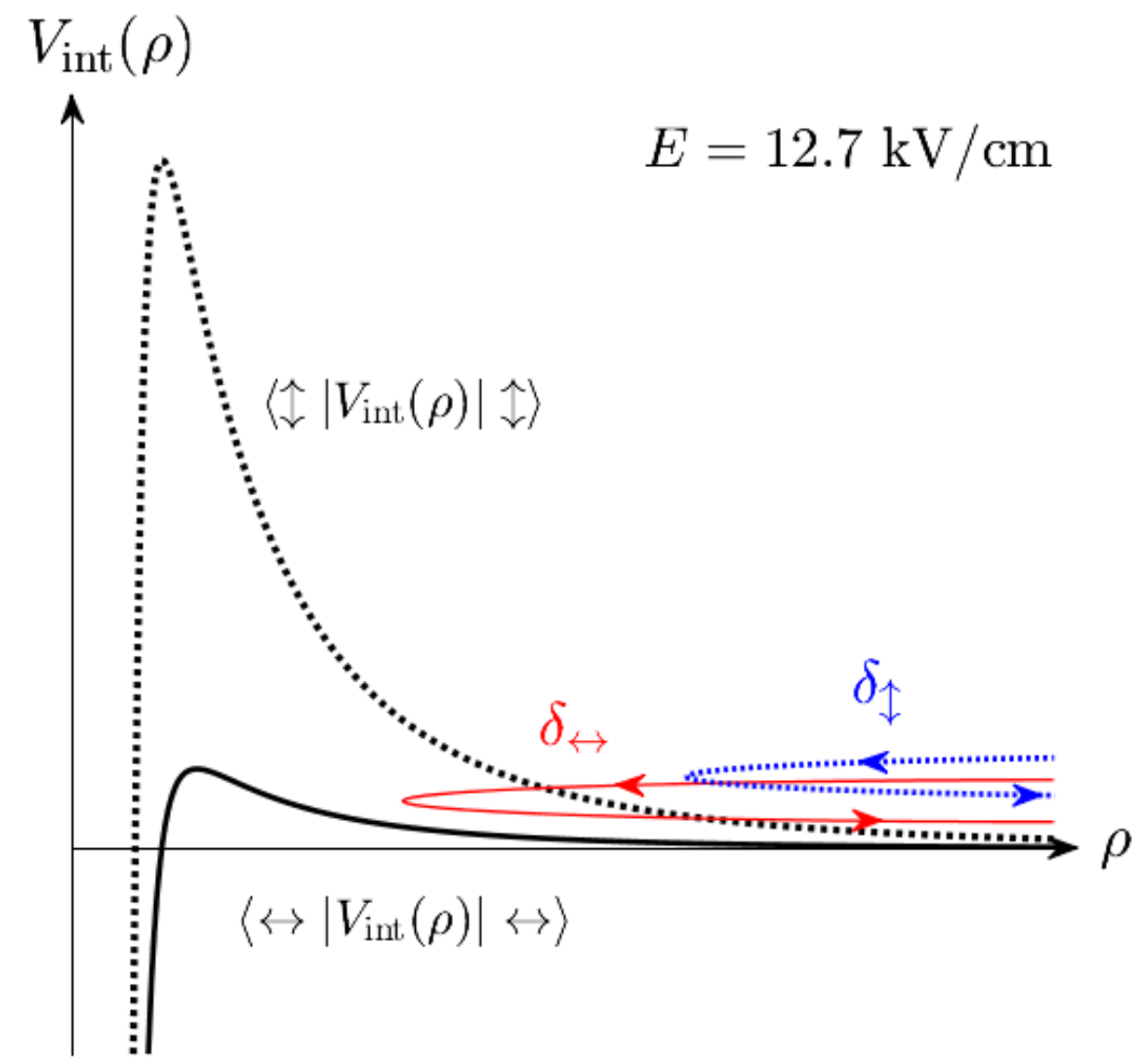

FIG. S4. A schematic diagram of the differential phase shifts (S29) developed from scattering off the interaction potential $V_{\mathrm{int}}(\rho)$. $\rho$ denotes the distance between the 2 colliding molecules in 2D. As illustrated, the $|\updownarrow\rangle$ (dotted curve) and $|\leftrightarrow\rangle$ (solid curve) scattering channels adiabatically connect to different corresponding potential energy surfaces over which the molecules accrue their scattering phase shifts, $\delta_{\updownarrow}$ and $\delta_{\leftrightarrow}$ respectively. The colliding molecules enter each channel with the same collision energy, where the asymptotic threshold energies are ignored in this figure for ease of comparison.

### 1. Quasi-2D contrast decay

For the analysis in quasi-2D, we utilize the forward scattering phase shifts read off from Eq. S16 as:

$$e^{2i\delta_\nu^{2D}(k)} \approx 1 - 2i(ka_D^\nu)\mathrm{Erfc}\,(ka_{\mathrm{ho}})\,e^{(ka_{\mathrm{ho}})^2}, \quad \nu = \Downarrow, \Uparrow, \leftrightarrow, \tag{S32a}$$

$$e^{2i\delta_-^{2D}(k)} \approx 1 - 2i(ka_D^-)\left(\frac{4}{3\sqrt{\pi}}\frac{1}{ka_{\mathrm{ho}}} - \mathrm{Erfc}(ka_{\mathrm{ho}})e^{(ka_{\mathrm{ho}})^2}\right). \tag{S32b}$$

Utilizing the backward scattering phase shift yields the same result in the derivation that follows. With an average 2D density of $n_{2D} \approx 1.2 \times 10^7$ cm$^{-2}$ and a temperature of $T_{\mathrm{avg}} = 300$ nK, the thermally averaged contrast decay rate $\langle\Gamma\rangle$ is computed as a function of the electric field $E$, showing the trend in Fig. S5. Close to the Heisenberg point, we see from Eq. (S29) that the change in contrast scales with $\chi$ as $\Delta C \sim \chi^2$. The scaling of $\beta_{\mathrm{el}}$ with $\chi$ is more intricate, but if taken to be some polynomial in $\chi$, might have its lowest non-trivial term just be $\sim \chi$. Under this naive assumption, we fit a $|\chi|^3$ monomial to $\langle\Gamma\rangle$ that yields the red-dashed curve in subplot (b). We find that although the proposed $|\chi|^3$ dependence is appropriate to the current experiment, it is sensitively dependent on the harmonic oscillator length along the tightly confined axis, $a_{\mathrm{ho}}$. For smaller $a_{\mathrm{ho}}$ (tighter confinement), $\langle\Gamma\rangle$ follows a trend closer to $\chi^2$, whereas larger $a_{\mathrm{ho}}$ (weaker confinement) follows $|\chi|^3$ more strongly. This variation is illustrated in Fig. S6, where we fit a monomial in $\chi$ (dashed red line) to the theory-predicted contrast decay rates (black solid line) while floating the exponent on, and coefficient multiplying $\chi$.

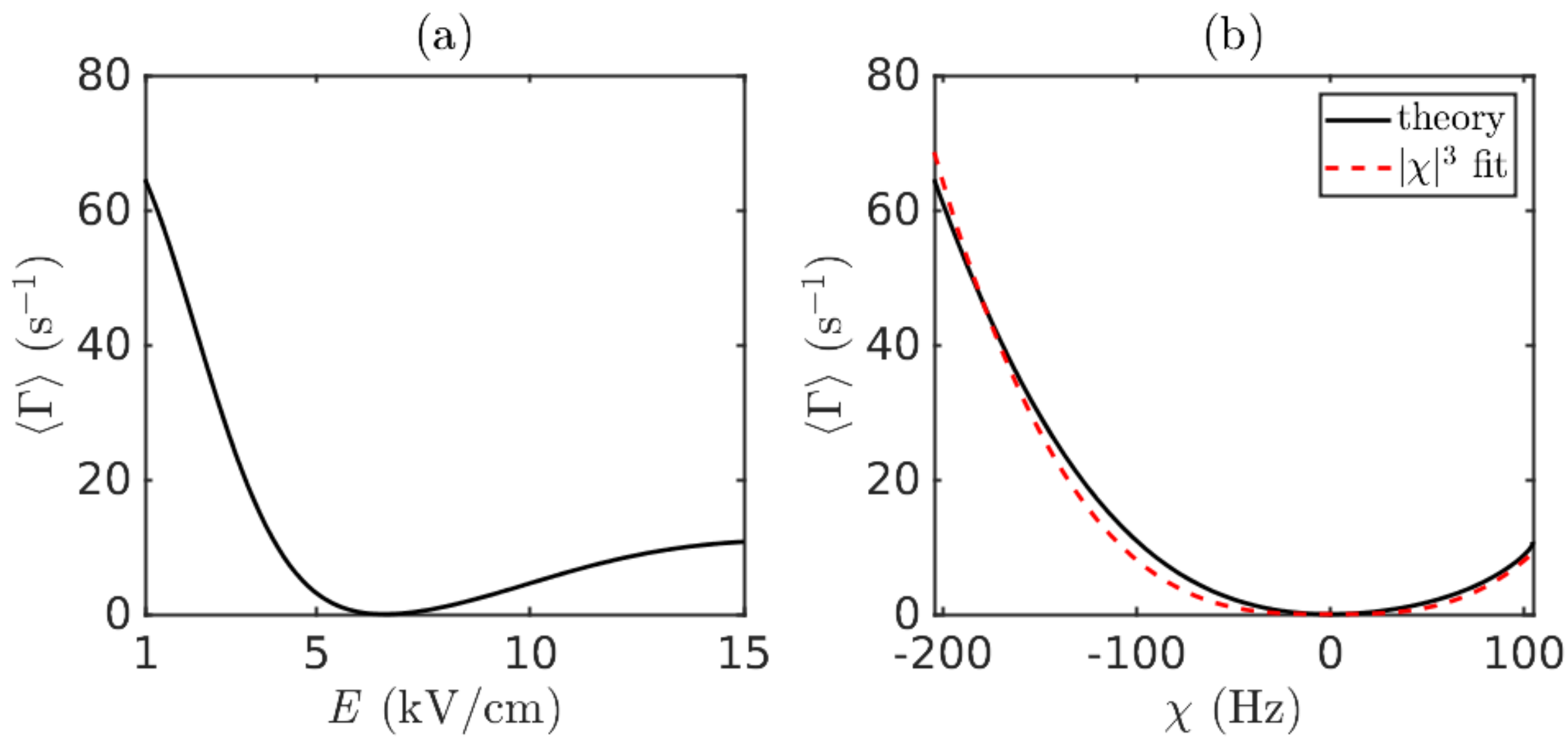

FIG. S5. The thermally averaged collisional contrast decay rate $\langle\Gamma\rangle$ in quasi-2D, vs (a) the electric field $E$ and (b) the interaction parameter $\chi$, in a quasi-2D geometry. A fit of $\langle\Gamma\rangle$ to a $|\chi|^3$ monomial is also included as the red dashed curve. The temperature is assumed to be $T_{\mathrm{avg}} = 300$ nK.

In reality, involvement of the antisymmetric spin sector in subsequent collisions will suppress contrast decay at long times (see Fig. S3 of appendix Sec. I A 1). This suppression is primarily due to 2-body losses, since molecules that would have otherwise decohered are now lost from the trap and do not contribute to the Ramsey measurement. The result is the observed stretched exponential behavior, which therefore renders the analytic theory developed here mostly qualitative.

### 2. Quasi-1D contrast decay

If the collisions occur in quasi-1D instead, a similar analytic derivation of the contrast decay rate can be performed utilizing the phase shifts and scattering rates in 1D. To do so, we must once again derive the Born approximated phase shifts in much the same way as we did in Sec. I A, with diagonal $S$-matrix elements

$$S_{\nu,\nu}^{1D}(k) = e^{2i\delta_\nu^{1D}(k)} = 1 - 2\pi i \langle k| \int d^2\boldsymbol{\rho} |\varphi_0(\boldsymbol{\rho})|^2 \langle\nu| \widehat{V}_{\mathrm{dd}}(\boldsymbol{r}) |\nu\rangle |k\rangle\,, \tag{S33}$$

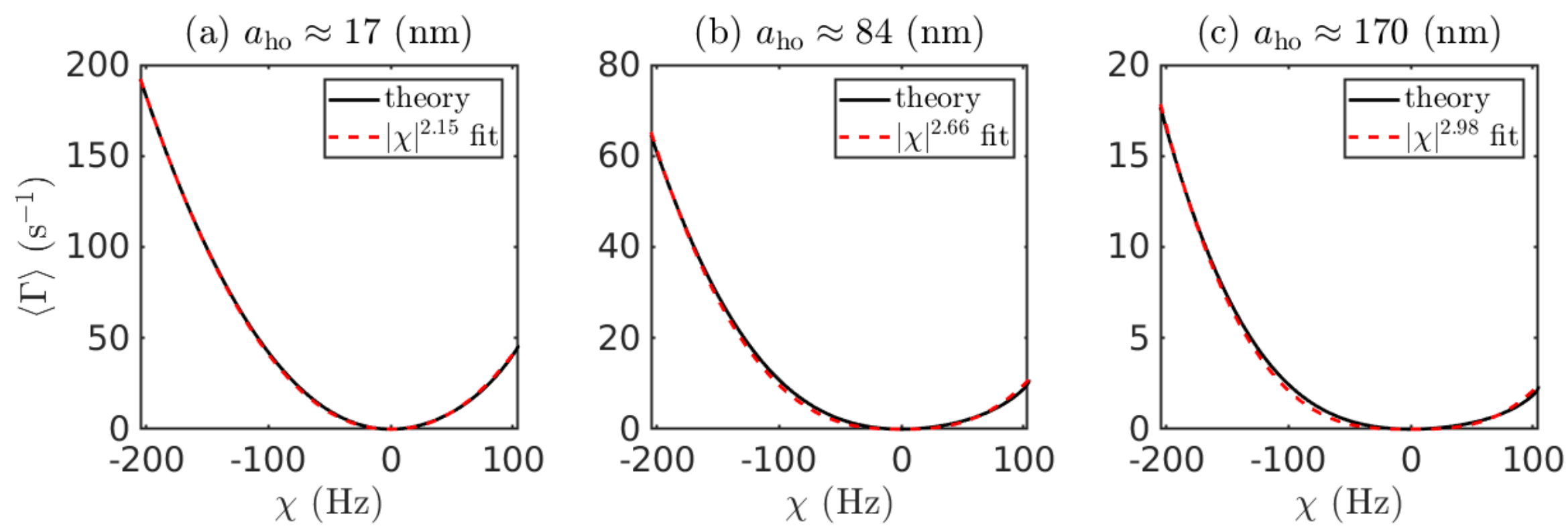

FIG. S6. The thermally averaged collisional contrast decay rate $\langle\Gamma\rangle$, vs the interaction parameter $\chi$ in a quasi-2D geometry, for 3 values of the harmonic oscillator length: (a) $a_{\rm ho} \approx 17$ nm, (b) $a_{\rm ho} \approx 84$ nm, (c) $a_{\rm ho} \approx 170$ nm. The temperature is assumed to be $T_{\rm avg} = 300$ nK.

where $\varphi_0(\boldsymbol{\rho})$ are the transverse harmonic oscillator ground states, along with the antisymmetrized ($\mathcal{A}$) and symmetrized ($\mathcal{S}$), energy-normalized momentum eigenstates:

$$|k\rangle_{\mathcal{A}} = i\sqrt{\frac{\mu}{\pi\hbar^2 k}}\sin(kx), \tag{S34a}$$

$$|k\rangle_{\mathcal{S}} = \sqrt{\frac{\mu}{\pi\hbar^2 k}}\cos(kx). \tag{S34b}$$

Evaluating the necessary integrals results in the quasi-1D elastic scattering rates (71) and phase shifts:

$$\beta_\nu^{\rm 1D}(k) = n_{\rm 1D}\frac{\hbar k}{\mu}\left|1 - e^{2i\delta_\nu^{\rm 1D}(k)}\right|^2, \tag{S35a}$$

$$e^{2i\delta_\nu^{\rm 1D}(k)} \approx 1 - 2i(ka_D^\nu){\rm E}_1\left(k^2 a_{\rm ho}^2\right)e^{(ka_{\rm ho})^2}, \tag{S35b}$$

where ${\rm E}_1(z) = \int_z^\infty dt' e^{-t'}/t'$ is the exponential integral. From the expressions above, we see that the low energy 1D elastic phase shift scales as $\delta(k) \sim k\ln(k)$, consistent with (72). For the symmetrized channels in which dipole-dipole interactions remain repulsive, the imaginary part of the complex phase shifts are exponentially suppressed (71), so we ignore them here as well. Then utilizing the thermally averaged contrast decay rate (S30) once more, a gas with a 1D density of $n_{\rm 1D} \approx 1.6 \times 10^3$ cm$^{-1}$ at $T_{\rm avg} = 300$ nK sees $\langle\Gamma\rangle$ is as a function of $E$ and $\chi$ as shown in Fig. S7.

## II. CORRECTIONS BEYOND THE $t$-$J$-$V$-$W$ MODEL

| Correction term | Two-body simulations | EMACE |
|---|---|---|
| next-nearest neighbor tunneling | ✔ | ✔ |
| full dispersion | ✔ | ✗ |
| external trap | 1D only | ✔ |
| higher bands | 1D only | ✗ |
| single-molecule loss | ✔ | ✗ |
| $s$-wave losses (on-site and nearest-neighbor) | ✔ | ✗ |
| on-site interactions (dipolar + contact) | ✔ | ✔ |
| corrected off-site interactions | ✔ | ✔ |

TABLE S1. Corrections beyond the $t$-$J$-$V$-$W$ model included in the simulations. The checkmark "✔" signifies that the given correction was included. The cross "✗" signifies that the given correction was not included. "1D only" means that the correction was only taken into account in the 1D simulations, but not in 2D.

The $t$-$J$-$V$-$W$ model describes the fundamental competition between dipolar interactions and motion studied in the experiment. However, several features of the experiment can lead to corrections beyond the pure $t$-$J$-$V$-$W$ model and thus need to be taken into account in numerical simulations to reproduce the contrast dynamics. These corrections are summarized in Tab. S1 that shows which corrections were taken into account in each simulation.

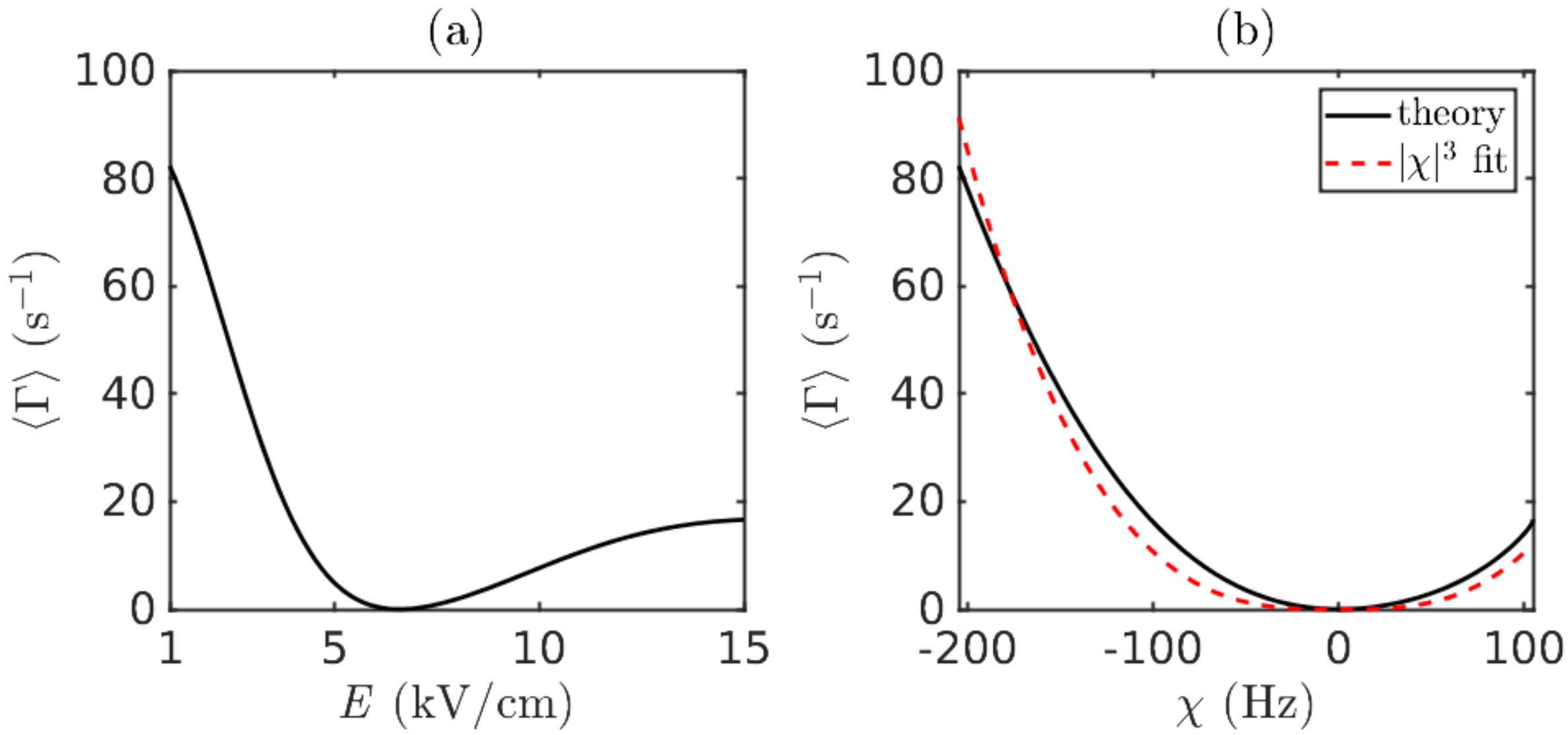

FIG. S7. The thermally averaged collisional contrast decay rate $\langle\Gamma\rangle$ in quasi-1D, vs (a) the electric field $E$ and (b) the interaction parameter $\chi$, in a quasi-1D geometry. A fit of $\langle\Gamma\rangle$ to a $|\chi|^3$ monomial is also included as the red dashed curve. The temperature is assumed to be $T_{\rm avg} = 300$ nK.

The first four rows describe corrections to the coherent single-molecule dynamics. While in deep lattices ($V_L \gtrsim 10E_r$), the single-molecule dynamics is well described by nearest-neighbor tunneling $t$ leading to a cosinusoidal dispersion relation, for shallower lattices this approximation breaks down. As a consequence, corrections need to be taken into account. In the two-body model, we work with single-molecule eigenstates, where the modified dispersion relation can be taken into account fully. In contrast, when working with localized orbitals in a tight-binding model, we need to include long-range tunneling of the form $t_\nu \hat{c}_i^\dagger \hat{c}_{i+\nu} + h.c.$ In practice, only nearest and next-nearest neighbor tunneling terms ($\nu \leq 2$) are included, since even in the extreme case, third-nearest neighbor tunneling is limited to $t_3 \approx 30$ Hz for $V_L = 0.5E_r$, when nearest neighbor tunneling is almost $t_1 \approx 300$ Hz and next-nearest neighbor tunneling is $t_2 \approx -70$ Hz. As discussed in the main text and below, the external trap is an important correction, especially in 1D, where it can lead to full localization of the molecules. Since it breaks translational invariance, it makes two-body simulations significantly more costly and is thus only included in the 1D simulations shown in Fig. S9(b) in the next section. For shallow lattices $V_L \lesssim 5E_r$, thermal population of motionally excited bands becomes non-negligible, enhancing delocalization and competing with the external trap. Higher bands population was therefore included in the two-body simulations for the 1D case where it competes with the trap. In previous EMACE simulations done without the harmonic trap and with slightly modified sampling, we found only minor corrections due to higher bands, so they have been excluded in all other cases for numerical efficiency.

In shallow lattices, molecule loss can slow down contrast decay due to a combination of selective loss of decohered molecules and dynamically decreasing density (see also Sec. I A 2). We include single-body and two-body $s$-wave loss in the two-body simulations with a non-hermitian Hamiltonian approximation but neglect it in the EMACE for simplicity. The $s$-wave loss is computed both on-site and for adjacent lattice sites.

Both two-body calculations and EMACE simulate the full dipolar Fermi-Hubbard model and thus allow two molecules to occupy the same lattice site. The on-site interactions are then computed by a combination of dipolar and contact interactions, where dipolar interactions typically dominate. Furthermore, off-site interactions are modified by the delocalization of Wannier orbitals, allowing molecules to approach each other more closely than point dipoles trapped in the lattice minima, as derived by Ref. (73). This can lead to a significant increase in interaction strength along the shallow lattice direction by more than a factor of 2. The corrections from the delocalization of two interacting molecules arise due to a combination of direct (diagonal part of interactions) and exchange (off-diagonal part, i.e. switch orbitals ) interaction processes . In practice, the exchange processes always have a much smaller rate than the direct ones and are thus neglected (see also (73)). We include corrections to interactions in a rectangular cuboid up to distance (3,1,1) for 1D and (2,1,2) for 2D and use the point-dipole approximation beyond that.

## III. CONFINING POTENTIAL

To understand the role of the confining potential, we can examine its effects in the single-band tight-binding model, where the single-particle Hamiltonian takes the form

$$\hat{H}_{\mathrm{sp},1} = -t \sum_{\sigma=\uparrow,\downarrow} \sum_{\langle i,j \rangle} \hat{c}^{\dagger}_{i,\sigma} \hat{c}_{j,\sigma} + \sum_i \mu_i \hat{n}_i \,, \tag{S36}$$

where the chemical potential $\mu_i = m \sum_\alpha \omega^2_{\mathrm{ho},\alpha} r^2_{i,\alpha}/2$ incorporates the effects of the confining potential, with $r_{i,\alpha}$ taken to be the spatial center of lattice site $i$. When $V_{L,x} = 0$, the single-particle eigenstates are approximately localized to a few lattice sites on the trap edges, whereas single-particle eigenstates towards the trap center are delocalized, and take the form of discretized harmonic oscillator states (51).

This behavior can be largely understood by restrictions on the available lattice sites a particle can tunnel to, owing to differences in the on-site energies that exceed the available tunneling bandwidth. That is, for nearest-neighbor tunneling rate $t$, a particle initially at site $i$ can only tunnel to sites $j$ for which $4t \lesssim |\mu_i - \mu_j|$. Sufficiently large differences on the order of the bandwidth or the addition of many-body processes can foster delocalization beyond this single-particle picture. Nonetheless, we expect this picture to provide a qualitative reference for the observed physics in the main text, at least away from shallow lattice depths where the effects of higher bands cannot be dismissed. Here, for example, trap-induced coupling to higher motional bands can help delocalize the particles in the limit $V_{L,x} \to 0$.

Within this picture, the separation between localized and delocalized single-particle eigenstates occurs at distances $r_{\mathrm{loc}} \approx \sqrt{4ht/m\omega^2_{\mathrm{ho},\alpha}}$ from the center of the confining potential (51). Even for lattices as shallow as $V_{L,x} = 5E_R$, this sets the localization radius to $\sim 8$ lattice sites. When considering the initial, 3D Gaussian distribution of the particles, this results in $\sim 17\%$ of the particles occupying single-particle eigenstates that are delocalized along the $x$ direction. When tunneling is enabled in both the $x$ and $y$ dimensions, an even smaller fraction, $\lesssim 3\%$ of the particles occupy the single-particle eigenstates that are delocalized in two dimensions at the trap center. Instead, as illustrated in the main text, most particles occupy single-particle eigenstates that are delocalized along quasi-1D rings of relatively fixed chemical potential energy that are a constant distance from the trap center. When tunneling is restricted to one dimension, the majority of the particles in the system remain fully localized at the trap center.

## IV. TWO-BODY SIMULATION

In this section, we describe a method to simulate the contrast decay dynamics in shallow lattices $V_{x,y,z} \lesssim 10E_r$, for which single molecule dynamics dominate over interactions and molecules delocalize quickly. In this case, it is convenient to describe dynamics in terms of single-molecule eigenstates, i.e. Bloch waves in the absence of a confining potential, or more complex states if the dipole trap is taken into account. The resulting physics is very similar to the collisional physics discussed in Sec. I, motivating a numerical method based on two-particle physics. Analogous to the free-space scattering simulation, we always consider pairs of molecules in isolation without feedback between different pairs' dynamics. We then extract the contrast of a single molecule by multiplying the effect of different molecules. The contrast of the full sample is then computed as the average over all molecules.

Since we work directly with single-molecule eigenstates, this approach allows us to treat shallow lattices even for fully delocalized molecules in 2D. It further allows us to distinguish the effects of mode-preserving scattering which leave the motional state unchanged and mode-changing collisions. This supports the scattering intuition for the shallow lattice contrast dynamics since mode-changing collisions dominate.

### A. Method

We start by separating the Hamiltonian into two-body terms

$$\hat{H} = \sum_{i>j} \hat{H}_{ij} \,, \tag{S37}$$

where $i$ and $j$ sum over all molecules. This requires us to keep track of all molecules as if they were distinguishable, which is a reasonable approximation far from degeneracy.

We can then compute the contrast as

$$C(T) = \frac{2}{N}\sum_i \langle\psi_0|\exp\Big(\mathrm{i}\hat{H}T\Big)\hat{s}_i^X\exp\Big(-\mathrm{i}\hat{H}T\Big)|\psi_0\rangle \tag{S38}$$

$$\approx \frac{2}{N}\sum_i \langle\psi_0|\left[\prod_{j>j'}\exp\Big(\mathrm{i}\hat{H}_{ij'}T\Big)\right]\hat{s}_i^X\left[\prod_{k>k'}\exp\Big(-\mathrm{i}\hat{H}_{kk'}T\Big)\right]|\psi_0\rangle \tag{S39}$$

$$\approx \frac{2}{N}\sum_n \langle\psi_0|\left[\prod_{j}\exp\Big(\mathrm{i}\hat{H}_{ji}T\Big)\right]\hat{s}_i^X\left[\prod_{k}\exp\Big(-\mathrm{i}\hat{H}_{ki}T\Big)\right]|\psi_0\rangle \tag{S40}$$

$$\approx \frac{2}{N}\sum_i\prod_j \langle\psi_0|\exp\Big(\mathrm{i}\hat{H}_{ji}T\Big)\hat{s}_i^X\exp\Big(-\mathrm{i}\hat{H}_{ji}T\Big)|\psi_0\rangle \tag{S41}$$

Here, the first step is a trotterization, which has errors on the order of the commutators of the two-body Hamiltonians $\left[\hat{H}_{ij}T,\hat{H}_{kl}T\right]$, which are small for short times, or when the two terms approximately commute. In the second step, we commute all terms $\hat{H}_{jj'}$ with $j,j'\neq i$ with $\hat{s}_i^X$ and cancel them with their partner. This step is exact for the appropriate ordering of terms in step Eq. (S39). In the third step, we assume that correlations that are built up by $\hat{H}_{ji}$ cannot be recovered by evolution induced by interactions with other pairs. This follows the spirit of independent collisions introduced in Sec. I, where we trace out the other molecule after each collision. In this approximation, the expectation value can be computed for each pair independently.

Physically, this can be interpreted as follows: Two-particle interactions directly lead to contrast decay due to spin-motion coupling and coherent contrast oscillation. The contrast oscillations are out-of-phase for different molecule pairs so that they are also observed as contrast decay. In addition, two-particle interactions also build up correlations. These correlations lead to direct contrast decay which is also included in the model. We ignore however higher-order effects, where such built-up correlations modify interactions with a third molecule.

### B. Simulation and Losses

The approximate contrast dynamics can thus be computed by only computing two-particle dynamics, avoiding the normally exponential growth of Hilbert space with molecule number. We assume that each molecule is initially in an eigenstate of the single-particle Hamiltonian along the shallow lattice direction(s) and in a localized Wannier orbital in the deep lattice direction(s). We sample the molecules $i$ from the appropriate probability distribution. For each molecule $i$, we independently sample surrounding molecules $i'$. We then compute contrast evolution according to Eq. (S41).

When sampling the molecules $i'$, we only include those in nearby layers/chains, since far-away molecules barely affect the dynamics. In particular, for 2D we include molecules up to two layers away, and in 1D we include molecules up to a cutoff distance of $(\Delta y,\Delta z)\leq(3,3)$ independently in both directions. In 1D, we sample molecule pairs from the full eigenstate including the external trap and the first excited band according to a thermal probability distribution. In 2D, we ignore the dipole trap for simplicity and assume that molecules are evenly distributed among the lowest band. This speeds up calculations by including quasi-momentum conservation. In 2D, we then estimate the full contrast dynamics by a local density approximation, i.e. we weigh the dynamics computed at a given density with a density resolution of 1% with the fraction of the number of molecules at that density.

To estimate the effect of losses, we include an imaginary $s$-wave scattering component describing chemical reactions $\mathrm{Im}(a_s)\approx 700a_0$ and a single molecule loss rate $\gamma_1 = 2.5\mathrm{s}^{-1}$, both chosen to approximately match the experimentally observed loss. We then compute the singlet channel dynamics with the non-hermitian Hamiltonian. The norm decay of this dynamics is a measure of the losses (in the singlet channel). We estimate the total collisional loss rate of molecule $i$ with all other molecules $j$, $\gamma_2^{(i)}$ from the short-time dynamics between 1ms and 10ms chosen to avoid fast short-time decay due to initial doubly occupied sites and long-time saturation of the population. This allows us to approximate the population time evolution of molecule $i$ as

$$p_i(T+\Delta T) = p_i(T)\exp\Big(-\gamma_1\Delta T - \gamma_2^{(i)}p_i(T)P_S(T)T\Big)\,, \tag{S42}$$

with $P_S(T)=\frac{1-C_{i,\mathrm{bare}}(T)^2}{4}$ the probability to find one pair in the singlet state. Here, we assume that $p_i(T)$ can also be used to approximate the remaining population of the other molecules. When computing the contrast, we further assume that $s_i^X(T)$ also decays with single molecule decay rate: $s_i^X(T) = s_i^{X(0)}(T) \times \exp(-\gamma_1 T)$, where $s_i^{X(0)}$ is the contrast computed without including losses.

Finally, we include feedback of the loss into the contrast decay, which decays slower for small particle number, as

$$C_{i,\text{corrected}}(T+\Delta T) = C_{i,\text{corrected}}(T) \times \left[\frac{C_{i,\text{bare}}(T+\Delta T)}{C_{i,\text{bare}}(T)}\right]^{p_i(T)} . \tag{S43}$$

Here, $C_{i,\text{bare}}$ is the contrast computed for molecule $i$ [Eq. (S41) without the sum].

### C. Regime of validity of the two-body model

This simulation is expected to work well in shallow lattices with low filling fractions. For these systems, it is unlikely that more than two particles approach each other closely at any given time, and the physics is dominated by individual two-particle collisions, which are here computed in real-time. We note that the trotterization and truncation to two-body terms made above become exact for short times and for an Ising model of frozen spins, respectively.

One term that is neglected in this model can be understood by considering three particles $A$, $B$, and $C$ in a deep lattice with exchange interactions. Let us focus on $\langle \hat{s}_A^X \rangle$. Then, as discussed above, the interaction $\hat{H}_{AB}$ builds up correlations of the form $\hat{s}_A^Y \hat{s}_B^Z$. This correlation can then be transferred to a correlation between molecules $A$ and $C$ by $\hat{H}_{BC}$. Finally, this will feed back into a contrast measurement through $\hat{H}_{AC}$. This type of interaction is especially relevant in a deep lattice, where particles $A$, $B$, and $C$ stay localized and thus continue to interact strongly with one another. As a consequence, the two-body approximation should stay valid here only in the Ising case or at short times.

In the shallow lattice, the picture becomes more complicated: Now, correlations build up not only between the spin degrees of freedom of different molecules but also between spin and motional degrees of freedom of the same or different molecules. However, any entanglement between internal and external degrees of freedom is typically only observed as decoherence of the contrast, and as such should be sufficiently captured by the two-particle model. In addition, itinerant molecules can now move apart after interactions, such that circular interactions between three particles as described above become increasingly rare as molecules $A$ and $B$ separate, such that it is unlikely that molecule $C$ interacts with both of them. However, for finite systems, at long times particles can explore the entire system, such that the three-particle physics described above becomes relevant again. This is especially relevant in 1D, where the spatial domain that particles can explore before returning to their initial position is relatively small.

### D. Mode-changing and mode-preserving collisions

We can use these two-particle simulations to shed light on the difference between mode-changing and mode-preserving collisions. In particular, two distinct mechanisms can lead to contrast decay. In the absence of collisions that change the external state, e.g. because such collisions are energetically suppressed, molecules remain in their initial single-molecule eigenstates. In this case, the coherent dipolar interactions in Eq. (1) of the main text induce coherent spin dynamics between dipoles pinned in mode-space. The resulting model can be written as a mode-space spin-model (38) with density-density, density-spin, and spin-spin interactions. These interactions, in particular the differential between spin-Ising and spin-exchange interaction, set the time scale of dephasing, analogous to the very deep lattice case.

On the other hand, if interactions are strong enough to couple different motional eigenstates, this leads to collisional decoherence. E.g. in 2D without transversal lattice, there are always energetically allowed lateral collisions which conserve the magnitude of relative momentum $k$, but change its direction. In the Born approximation, the collision rate is proportional to (as discussed in Sec. I)

$$\beta(k) = n\sigma(k)v(k) \propto n\langle \hat{H}_{\text{dd}} \rangle_k^2 \rho_E(E_k)^2 v(k) \,. \tag{S44}$$

Such collisions then lead to an entanglement of spin and motional degrees of freedom, which leads to irreversible contrast decay. Notably, all parameters except $n$ can be modified by the lattice.

To quantify the relative importance of both processes, we simulate both the full Hamiltonian and the spin model, only. The results are shown in Figs. S8. For all parameters, the spin-model contrast is longer lived. Interestingly, at both weak and strong $\boldsymbol{E}$-fields, the contrast decay predicted by the spin model almost is only slightly faster at the shallowest lattices of $V_x = V_z = 0.5E_r$. As the lattice depth is increased, the deviation between both descriptions increases. Strikingly, at 1 kV/cm, this even leads to a reversal of the order of contrast decay rates: While within the spin model contrast decay is predicted to be slowest in the $7E_R$ lattice, contrast decay is actually slowest at $0.5E_R$ when taking all terms into account, consistent with the experimental observation. The increased importance of mode-changing collisions as lattice depth is increased from zero is attributed to higher density of states and slower velocity, as can be seen in Eq. (S44).

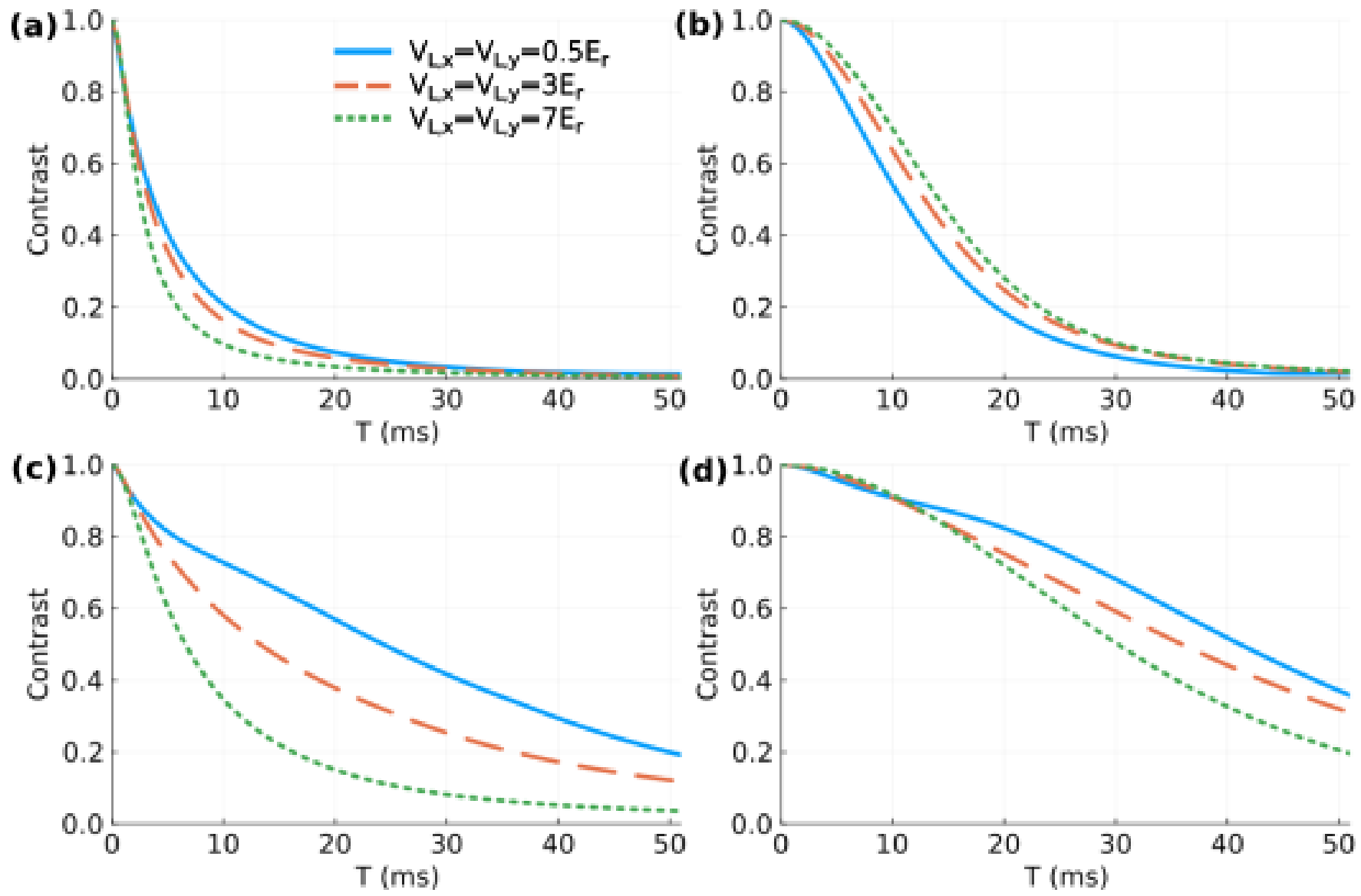

FIG. S8. Contrast dynamics computed from the two-body model at $|\boldsymbol{E}| = 1\text{kV/cm}$. (a) Simulation of all terms including mode-changing collisions. (b) Simulation of only spin-model terms. Both simulations are for $N_{\text{mol}} = 10,000$ molecules, computed in a local density approximation and with a $y$-lattice depth of $65E_R$. Differently colored and patterned traces correspond to traces at different lattice depths $V_x = V_z$ in $x$- and $z$-direction. (c), (d) Same as (a), (b) for $|\boldsymbol{E}| = 12.7\text{kV/cm}$. Simulations in a $20 \times 20$ lattice.

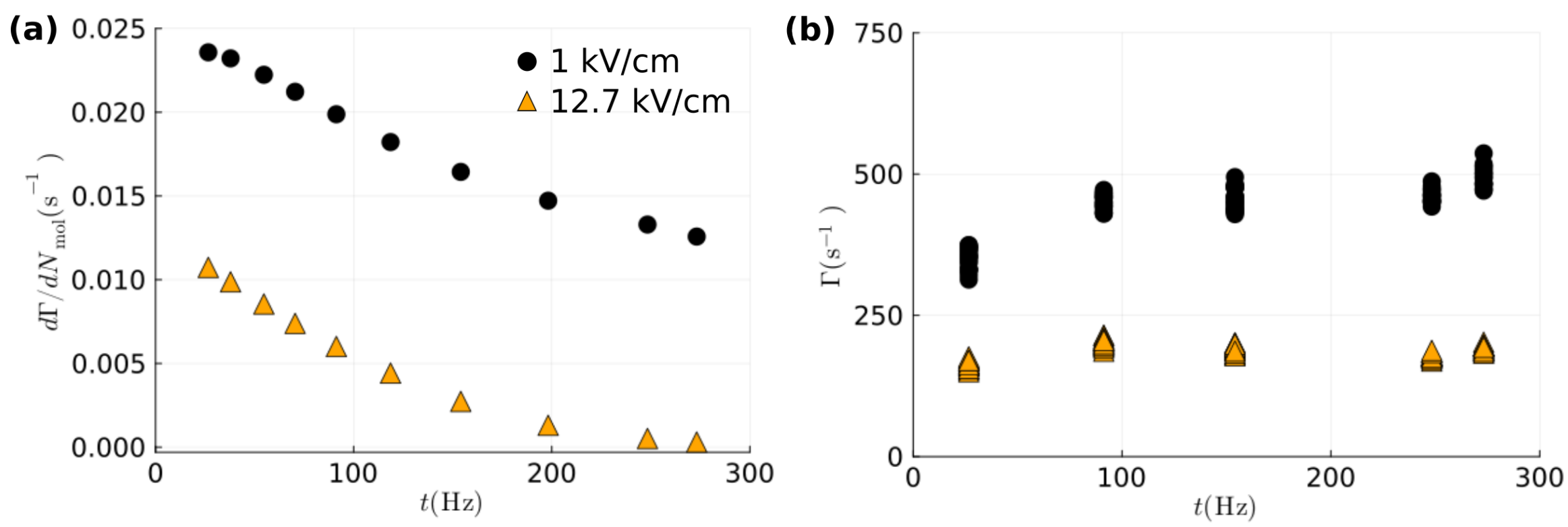

FIG. S9. Decay rates predicted by the two-body model. (a) Slope of the contrast decay rate vs molecule number (proportional to 2D density) for different lattice depth $V_{L,x} = V_{L,z} \leq 10E_r$ in the 2D tunneling scenario. As for experimental data, contrast decay rates were extracted by a stretched exponential fit. Black circles are in the exchange-dominated case ($|\boldsymbol{E}| = 1$ kV/cm), while orange triangles are in the Ising-dominated case ($|\boldsymbol{E}| = 12.7$ kV/cm). The slope was calculated for molecule numbers $2,500 < N_{\text{mol}} < 15,000$ in a local density approximation. Each sample is computed for $25 \times 25$ lattice sites. (b) Contrast decay rates in 1D extracted from a stretched exponential fit to the dynamics. Sample temperature $T_{\text{avg}} = 250$ nK. We include 61 lattice sites in the shallow direction, with an average of 6 molecules per chain. Different points are bootstrapped results from different disorder realizations.

### E. Results of the two-body model

We now summarize the numerical results obtained for different lattice depths, $\boldsymbol{E}$-fields, and molecule numbers in a way similar to the experimental data: We first fit each numerical time trace with a stretched exponential. We then fit these decay rates as a function of molecule number $N_{\mathrm{mol}}$ with a linear function. This slope is proportional to the experimental quantity $\kappa$, which uses the 2D density instead of the total number of molecules. Fig. S9(a) shows the fitted slope as a function of $t$ for the two different $\boldsymbol{E}$-fields. We only show shallow lattices with depth less than $10E_r$, since the simulations assume that the single-molecule eigenstates form a good basis, which is not true in deep lattices, where bands get narrow and the interaction energy can be much larger than the effective kinetic energy due to the band structure. For both electric fields, there is a significant reduction in $d\Gamma/dN_{\mathrm{mol}}$ with decreasing lattice depth, i.e. increasing $t$.

We now turn to the 1D scenario in Fig. S9(b). As discussed in the main manuscript, the external trap here leads to localization of molecules in all 3 dimensions and thus cannot be ignored. We thus compute single molecule eigenstates including both the lattice and the optical dipole trap. Here, we only show the decay rate at a fixed density of 10%, which we expect to behave similarly to the density-normalized decay rate $\kappa$. In 1D in the presence of an external trap, we find a strong dependence of the fitted decay rates on the specific positions of all molecules. We compute the contrast dynamics for 300 molecules in a given environment. We then use bootstrapping to compute average trajectories and decay rates. The different symbols correspond to 20 bootstrapping samples. As in the experiment, the simulated decay rate at 1 kV/cm is significantly higher than that at 12.7 kV/cm. The rate at 12.7 kV/cm has no significant dependence on the tunneling rate. In contrast, and opposite to the scaling in 2D, at 1 kV/cm, the contrast decay rate *increases* with decreasing lattice depth. This is qualitatively consistent with experimental observations, which show the disappearance of the peak as a function of the tunneling rate in 1D.

## V. EFFECT OF DENSITY-DENSITY INTERACTIONS

To understand the effect of the density-density interaction $V$, we simulate the dynamics for the true system parameters, and for manually setting $V = 0$ and $V = 100$ Hz. As can be seen in Fig. S10, the dynamics with the actual values of $V$ are virtually indistinguishable from the dynamics for $V = 0$, suggesting that it is unlikely that the experimentally observed dynamics are arising from density-density interactions. Even when we artificially set $V = 100$ Hz in the two-body simulations, the shallow lattice dynamics is barely modified with respect to the actual $V$. For intermediate depth lattices of $10E_R$, however, a very large $V$ could stabilize the contrast.

We attribute the relative insensitivity of the contrast dynamics to $V$ to the fact that $V$ only directly depends on the local density and not on the spin. As a consequence, $V$ can only coherently modify local densities, which can only indirectly feed back into the contrast dynamics. In the very deep lattice, where there is no coherent mechanism for population change, the $V$ term thus becomes a constant and cannot affect any dynamics. For very shallow lattices, in contrast, where $t > V$, density interactions are not strong enough to significantly affect motional dynamics. Only in the intermediate regime, when $0 \ll t < V$ could there be an effect of $V$. In the extreme case, one might consider that strong repulsive density-density interactions force the molecules to partially localize, thus removing disorder from the system and enhancing the contrast. This would however require interactions that are both much larger than tunneling and than the sample temperature. Since $V < 20$ Hz remains small for all experimental parameters [panel (a)], we thus conclude that $V$ does not significantly modify the contrast dynamics.

## VI. EXTENDED MOVING AVERAGE CLUSTER EXPANSION (EMACE)

### A. Moving average cluster expansion (MACE) for stationary dipoles

Here, we describe a method to compute the relaxation dynamics of the $t$-$J$-$V$-$W$ model, inspired by the moving-average cluster expansion (MACE) for a system of randomly distributed dipoles (42). First, we introduce MACE applied to a sparsely filled lattice of stationary molecules. In essence, this method consists of solving the dynamics of each dipole in the presence of its most strongly coupled neighbors, forming a local *cluster* of particles. The dynamics of global observables are then averaged over all possible clusters in the system. Let $i_a$ denote the indices of lattice sites initially populated with a molecule, where $a = 1, ..., N_{\mathrm{mol}}$ labels the particle index. We assume each of these sites to be populated by a single molecule in the $(|\uparrow\rangle + |\downarrow\rangle)/\sqrt{2}$ state. We then form a cluster of $M$ lattice sites, composed of site $i_a$ in addition to occupied sites $i_{a'}$ with the $M-1$ largest absolute couplings $|V_{i_a,i'_a}|$ to $i_a$; that is, we have sets

$$\mathcal{C}_M(i_a) = \{i_a, i_{a_1}, ..., i_{a_{M-1}}\} \tag{S45}$$

such that $|V_{i_a,i_{a_1}}| \geq |V_{i_a,i_{a_2}}| \geq ... \geq |V_{i_a,i_{a_{M-1}}}| \geq |V_{i_a,i_{a'}}|$ for all $i_{a'} \notin \mathcal{C}_M(i_a)$. We refer to $i_a$ as the *nucleus* of cluster $\mathcal{C}_M(i_a)$.

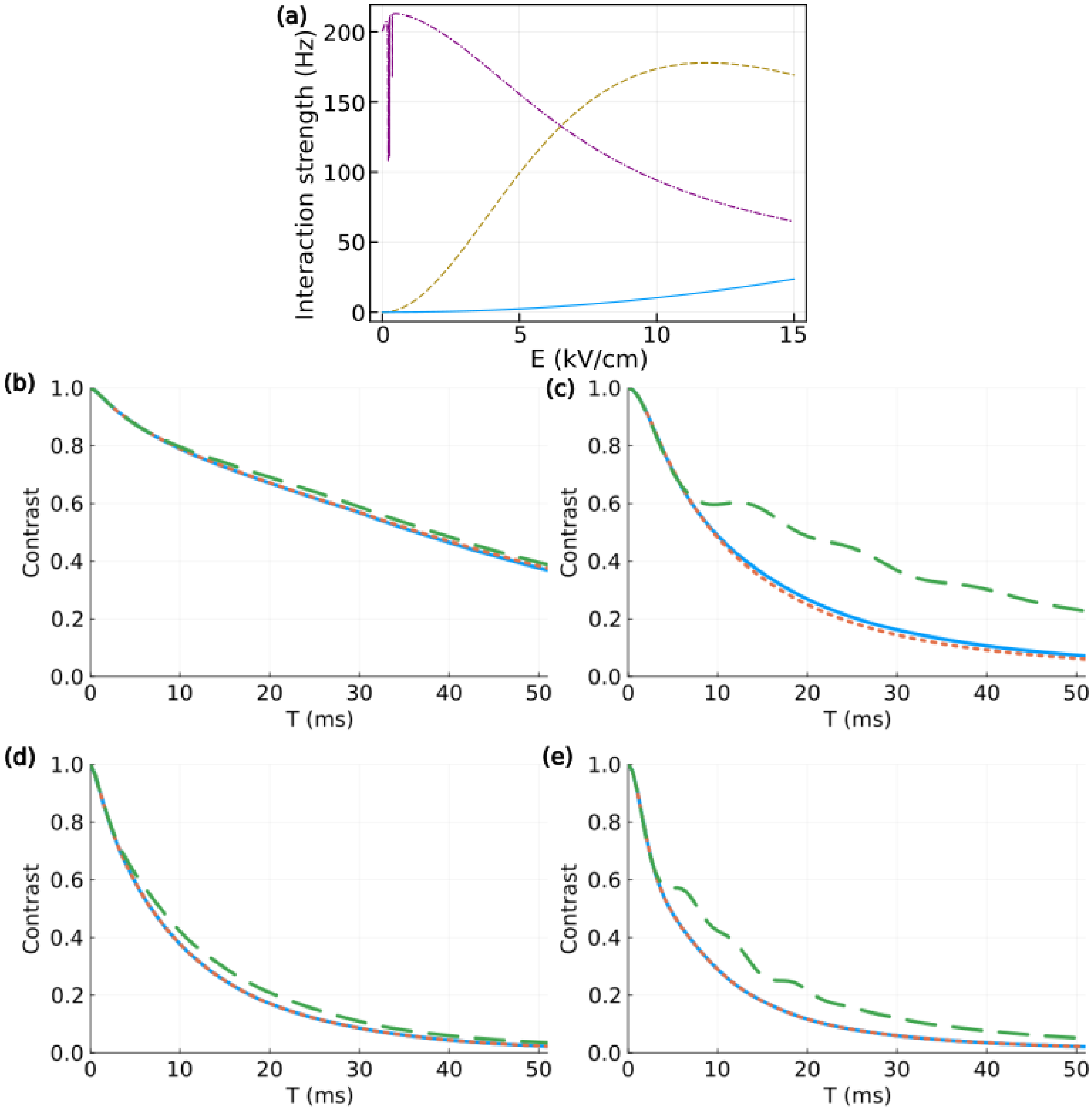

FIG. S10. Effect of density-density interactions $V$ in the 2D geometry. (a) Magnitude of the density-density interactions $V$ (blue solid) compared to $J_\perp$ (dashed dotted purple) and $J_Z$ (dashed gold) as a function of $|\boldsymbol{E}|$. (b-e) Contrast decay dynamics predicted by the two-body model for $V$ given by KRb parameters (blue solid), for $V = 0$ (orange dotted) $V$ and for $V = 100$ Hz (green dashed). Panels are for (b) $|\boldsymbol{E}| = 12.7$ kV/cm, $V_{L,x} = V_{L,z} = 1E_R$; (c) $|\boldsymbol{E}| = 12.7$ kV/cm, $V_{L,x} = V_{L,z} = 10E_R$; (d) $|\boldsymbol{E}| = 1$ kV/cm, $V_{L,x} = V_{L,z} = 1E_R$; (e) $|\boldsymbol{E}| = 1$ kV/cm, $V_{L,x} = V_{L,z} = 10E_R$. Simulations for $N_{\mathrm{mol}} = 5000$ in a local density approximation. For each local density, we simulate dynamics in a homogeneous $25 \times 25$ lattice.

The local Hilbert space for this cluster is given by

$$\mathcal{H}[\mathcal{C}_M(i_a)] = \bigotimes_{i \in \mathcal{C}_M(i_a)} \mathcal{H}_i, \tag{S46}$$

where $\mathcal{H}_i$ is the local Hilbert space corresponding to lattice site $i$. We may then solve for the unitary dynamics within this cluster:

$$|\psi_a(T)\rangle = \hat{U}_a(T)\, |\psi_a(0)\rangle \tag{S47}$$

where $\hat{U}_a(T) = \exp\{-i\hat{H}_a T\}$ and $\hat{H}_a$ is our Hamiltonian projected into $\mathcal{H}[\mathcal{C}_M(i_a)]$.

For computing local observables at the cluster nucleus $i_a$, e.g. $\langle \hat{O}_a(T) \rangle$, we utilize the values obtained from the corresponding cluster wavefunction $|\psi_a(T)\rangle$. We thus make the replacement $\langle \hat{O}_a(T) \rangle \approx \langle \psi_a(T)| \hat{O}_a |\psi_a(T)\rangle$; the contrast dynamics are then obtained via

$$C(T) \approx \frac{2}{N_{\mathrm{mol}}} \sum_{a=1}^{N_{\mathrm{mol}}} \langle \psi_a(T)| \hat{s}^X_{i_a} |\psi_a(T)\rangle \,. \tag{S48}$$

Note that, in principle, this is exact in the limit $M \to N_{\mathrm{mol}}$, where the cluster size approaches the total number of particles in the system. For $M < N_{\mathrm{mol}}$, we would expect deviations to occur on timescales no sooner than the inverse of largest neglected couplings, though how such deviations manifest is dependent on the observables of interest as well as on the expended range of the correlations that develop. As such, we expect that simple, low-order observables may even be relatively robust to deviations at longer times.

As in Ref. (42), we find that decent convergence of this observable over timescales relevant to the experiment can be achieved for clusters of size $M = 6$ for $\chi < 0$, which corresponds to the $J_\perp$-dominated regime; for $\chi > 0$, which corresponds to the $J_Z$-dominated regime, convergence can be achieved for clusters as small as $M = 2$. Throughout the main text, we always utilize a cluster size of $M = 8$. As noted in the main text, we observe an asymmetry in the dynamics for $\chi > 0$ and $\chi < 0$, which MACE similarly predicts (see Fig. S11). In fact, the relatively good convergence of the dynamics with $\chi > 0$ when only considering the pairwise dynamics between neighboring dipoles suggests a relatively local, few-body character to the dynamics in this region. On the other hand, the need to retain larger cluster sizes for $\chi > 0$ suggests that the dynamics in this regime are more many-body in nature. In fact, the difference difference between exchange dominated and Ising dominated dynamics for isotropic spin systems was pointed out in Refs. (46, 47), where the exchange dynamics can enforce a relatively collective behavior in the system, leading to scalable entanglement generation, as opposed to the local nature of the Ising dominated dynamics. For the present case, the anisotropic nature of the 3D dipolar interaction prevents the development of collective symmetry in the system; nonetheless, over clusters of spins over relatively short length scales can still exhibit a collective enhancement of their local dynamics. Indeed, via MACE, we observe in Fig. S11 a faint oscillatory character to the dynamics relative to the Ising-dominated regime.

### B. Extended cluster method for itinerant dipoles

We now develop an extended-MACE (EMACE) to describe the dynamics for a system of *non-stationary* dipoles. In this scheme, we solve for the $t$-$J$-$V$-$W$ dynamics of a dipole that is allowed to tunnel within a "buffer zone" of sites surrounding its initial position, while under the influence of the most strongly coupled neighbors *at each* lattice site. The other dipoles in this modified cluster are stationary, though in principle individual buffer zones may be constructed for each dipole in the cluster at the expense of added computational complexity.

Consider the adjacency matrix $\mathcal{A}$ that describes the links between nearest-neighbor sites on the lattice, so that $\mathcal{A}_{ij}$ is 1 (0) depending on whether lattice sites $i$ and $j$ are nearest-neighbors; the single-band tight-binding Hamiltonian matrix for a single particle with only nearest-neighbor hopping is then given by $-t\mathcal{A}$. For molecule $a$ initially at site $i_a$, we first form a *buffer zone* of lattice sites $i$ that are connected to $i_a$ by up to $r$ applications of $\mathcal{A}_{ij}$, i.e. the set

$$\mathcal{B}_r(i_a) = \left\{ i \mid (\mathcal{A}^{r'})_{i,i_a} = 1 \text{ for } r' \le r \right\}. \tag{S49}$$

For *each* index $i \in \mathcal{B}_r(i_a)$, we now consider a cluster $\mathcal{C}_M(i)$ (see Eq. (S45)) of site indices such that $\mathcal{C}_M(i)$ contains the $M-1$ *initially occupied* sites $i'_a$ with the largest absolute couplings $|V_{i'_a,i}|$ to lattice site $i$. We then form the composite cluster $\mathcal{C}_{M,r}(i_a) = \bigcup_{i \in \mathcal{B}_r(i_a)} \mathcal{C}_M(i)$, and consider the corresponding local Hilbert space

$$\mathcal{H}[\mathcal{C}_{M,r}(i_a)] = \bigotimes_{i \in \mathcal{C}_{M,r}(i_a)} \mathcal{H}_i. \tag{S50}$$

As in MACE, we solve for the unitary dynamics of this cluster, obtaining the cluster wavefunction $|\psi_a(T)\rangle = \hat{U}_a(T)\,|\psi_a(0)\rangle$, utilizing our Hamiltonian projected into $\mathcal{H}[\mathcal{C}_{M,r}(i_a)]$. In this way, the molecule initially at site $i_a$ is allowed to tunnel to any sites in the buffer zone $\mathcal{B}_r(i_a)$, while also subject to the influence of the surrounding dipoles most strongly coupled to those sites.

For computing the contrast, in general we must consider the entire ensemble at once, as the indistinguishable nature of the itinerant particles prevents us from computing the contrast associated with a specific initial molecule. However, given the sparse filling considered here in conjunction with the restricted nature of the buffer zone, we can associate the contribution to the contrast from each buffer zone as the contrast associated with any particles initially populating sites in $\mathcal{B}_r(i_a)$. Assuming that the total particle number $N_a$ of the buffer zone remains conserved, then we may compute the contrast of the system via

$$C(T) \approx \frac{2}{N_{\mathrm{mol}}} \sum_{a=1}^{N_{\mathrm{mol}}} \sum_{i \in \mathcal{B}_r(i_a)} \frac{1}{N_a} \langle \psi_a(T) | \, \hat{s}_i^X \, | \psi_a(T) \rangle \,, \tag{S51}$$

where the factor of $1/N_a$ prevents any over-counting associated with multiple molecules inhabiting $\mathcal{B}_r(i_a)$. We note that while the clusters are constructed using the nearest-neighbor adjacency matrix, the unitary dynamics within each cluster may in principle include an arbitrary dispersion, and we generally consider tunneling elements beyond nearest-neighbor, when relevant.

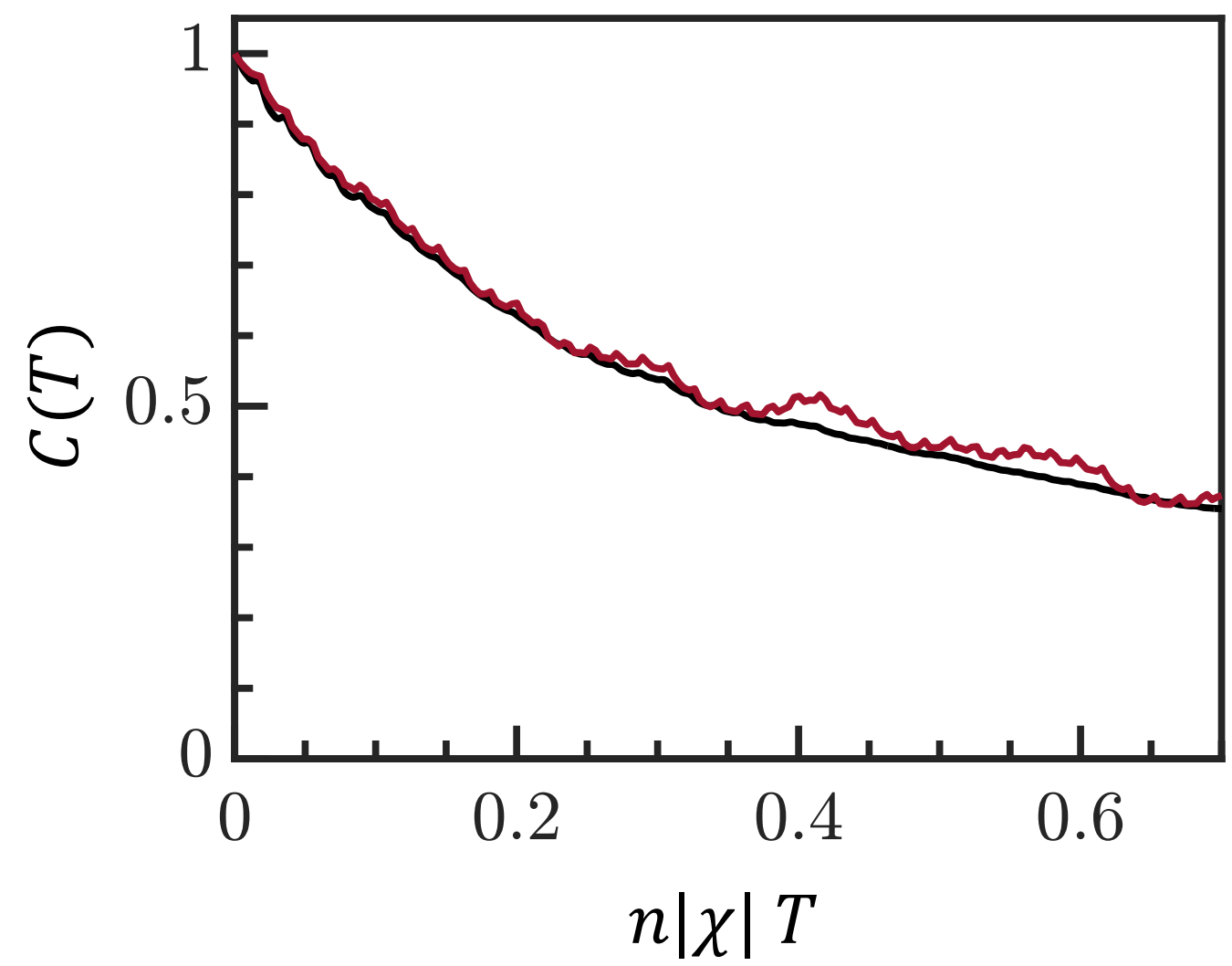

FIG. S11. MACE dynamics comparison for different $E$. MACE dynamics $E = 1.0$ kV/cm (red) and 12.7 kV/cm (black) for stationary dipoles. Time axis scaled by the interaction strength $|\chi|$ as well as the 3D particle density $n$ over considered system size. Results shown for $N_{\mathrm{mol}} = 500$ particles.

The surrounding dipoles are fixed in space, which constitutes an uncontrolled approximation in this scheme. In Fig. S12, we provide benchmarks of the contrast decay for the $t$-$J$-$V$-$W$ model with small numbers of particles, where we compare EMACE to the corresponding exact solutions. We choose a cluster size $M$ equal to the number of particles in the system, as well as choose $r$ sufficiently large so that the nucleus of each cluster is able to tunnel over all sites in the system connected to its initial location by the tunneling matrix. Thus, the only approximation made in EMACE is that the non-nuclear particles in each cluster remain stationary. We observe that, for the considered Hamiltonian parameters, there is essentially no difference between the contrast computed via exact diagonalization or via EMACE, providing substance to this approximation.

In practice, there are many additional modifications and subtleties we must make to the above scheme to enable feasible simulations of the system. First, the size the Hilbert space $\mathcal{H}[\mathcal{C}_{M,r}(i_a)]$ is typically still too large to solve via exact diagonalization methods except for very small $r$, and the long-ranged, 3D nature of the dipolar interactions makes tensor network approaches to the dynamics difficult. However, in the presence of an external confining potential, which we can model via an on-site chemical potential term in the Hamiltonian, $\sum_i \mu_i \hat{n}_i$, many lattice sites in the buffer zone may not be physically relevant for the corresponding single-particle dynamics, e.g. if the site differs too much its in potential energy from the initial site. When constructing buffer zones, we thus limit the allowed difference in tunneling energies between the initial site of the molecule and other sites in the buffer zone to some threshold, i.e. $|\mu_{i_a} - \mu_i| \gtrsim \Delta$. Our modified buffer zone than can be described via

$$\mathcal{B}_r(i_a; \Delta) = \left\{ i \mid (\mathcal{A}^{r'})_{i,i_a} = 1 \text{ for } r' \leq r;\, |\mu_{i_a} - \mu_i| \leq \Delta \right\}. \tag{S52}$$

If we examine only the single-particle eigenstates of a molecule tunneling in the presence of a confining potential, $\Delta = 4t$ is sufficiently large so that buffer zone does not exclude relevant sites that the particle might hop to. In the main text, we utilize $\Delta = 2t$ to further minimize the required computational resources. While this begins to artificially constrain the allowed sites to which a given dipole might hop to, it remains large enough so that the typical probability for a particle to occupy one of these neglected sites at late times remains $< 10\%$. Of course, many-body processes might further expand the allowed tunneling region, but we exclude this complexity here. We also do not consider the role of higher bands, which become relevant for sufficiently shallow lattices or sufficiently large variations in the site-to-site chemical potential. Lastly, we assume that the adjacency matrix element $\mathcal{A}_{ij}$ is strictly 0 for lattice sites along deep axial or deep radial directions.

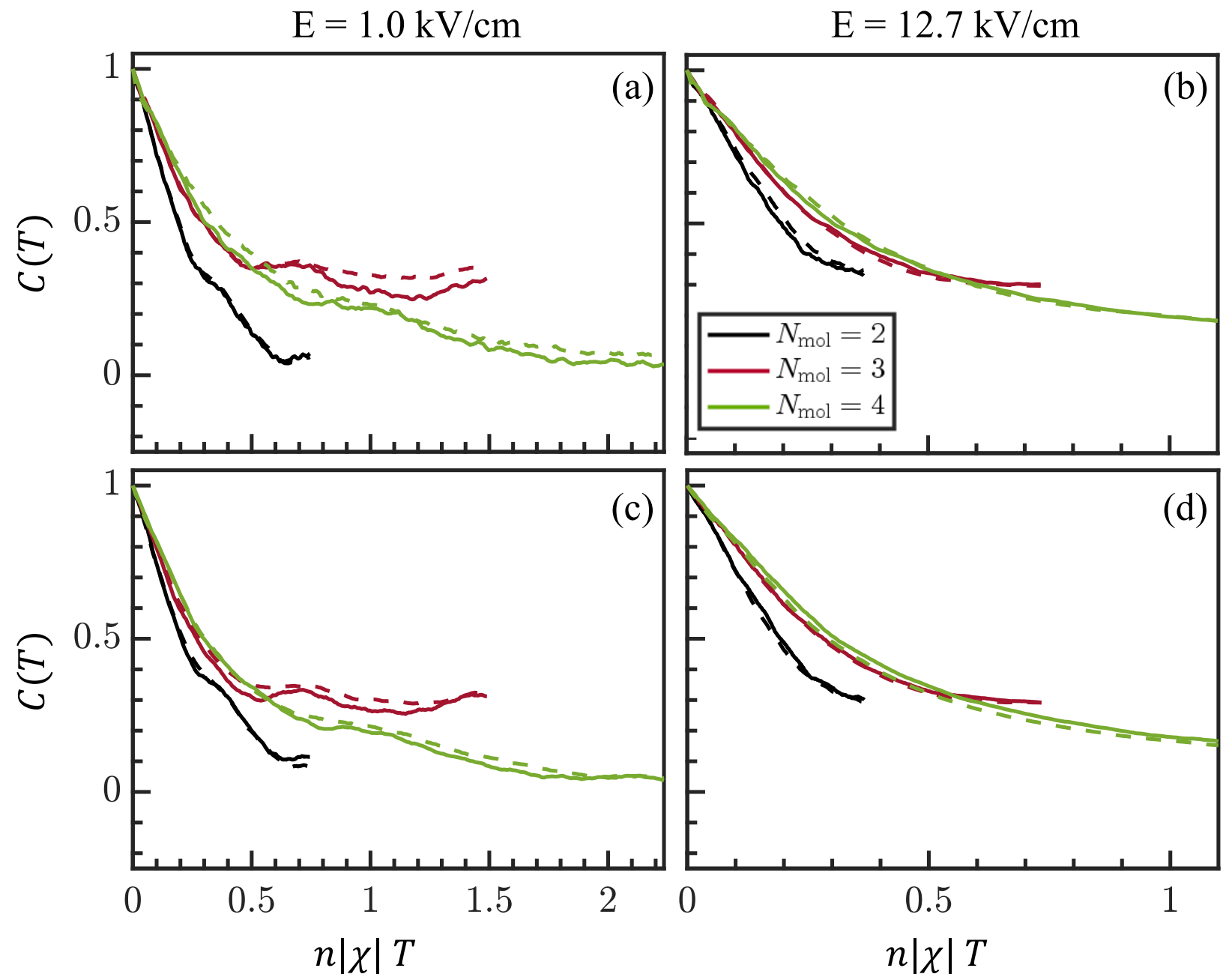

FIG. S12. Comparisons between EMACE and exact dynamics. (a,b) EMACE dynamics (solid) for varying small numbers of molecules $N_{\rm mol}$, compared to analogous exact dynamics (dashed). We simulate the system in an $L \times L \times L$ lattice with periodic boundary conditions with $L = 12$, where we average the results over $4 \times 10^3$ random initial configurations of the molecules, which are distributed uniformly over the system. Results are shown for tunneling along 1D in a $V_{L,x} = 1E_R$ lattice. We choose EMACE convergence parameters such that the buffer zone includes all available tunneling sites, and the cluster includes all particles in the system. We show results for $E = 1.0$ kV/cm (left) and 12.7 kV/cm (right). (c,d) Analogous results for $V_{L,x} = 10E_R$. Time axis scaled by the interaction strength $|\chi|$ as well as the 3D particle density $n$ over considered system size.

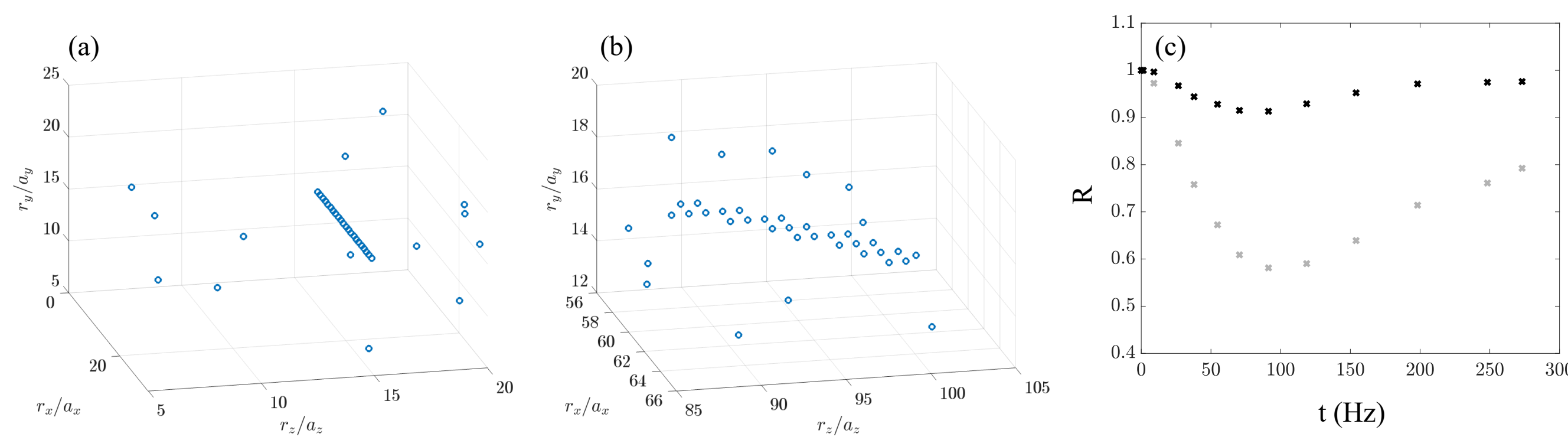

FIG. S13. Cluster formation in EMACE. Examples of lattice sites included in a typical cluster in the case of (a) 1D tunneling and (b) 2D tunneling, for a $5E_R$ lattice in each case. Buffer zones consist of large sections of contiguous lattice sites in each case. 1D example is taken near center of confining potential, where particles are relatively delocalized, and thus exhibit large buffer zones in the corresponding cluster. (c) Average figure of merit $R$ (see Eq. (S54)) for clusters constructed for EMACE at each lattice depth. Dark symbols denote values for EMACE convergence parameters used throughout the main text, with $\mathcal{N}_{\rm max} = 20$ and $M_{\rm max} = 15$. Lighter symbols denote values for a less stringent construction using $\mathcal{N}_{\rm max} = 30$ and $M_{\rm max} = 12$, resulting in a significantly lower figure of merit for most lattice depths.

The next modification we make is to limit the total amount of dipoles in each cluster to some value $M_{\rm max}$. For sufficiently large buffer zones, including the $M$ spins maximally coupled to each lattice site leads to untenable Hilbert space sizes. We thus exclude dipoles from this collection for which the mean coupling to the entire cluster is negligible, i.e. we form modified clusters

$$\mathcal{C}_{M,r}(i_a; M_{\max}) = \left\{i_a, i_{a_1}, ..., i_{a_{M_{\max}-1}}\right\} \subseteq \mathcal{C}_{M,r}(i_a) \tag{S53}$$

such that $|\sum_{i\in\mathcal{B}_r(i_a;\Delta)} V_{i,i_{a_1}}| \geq |\sum_{i\in\mathcal{B}_r(i_a;\Delta)} V_{i,i_{a_2}}| \geq ... \geq |\sum_{i\in\mathcal{B}_r(i_a;\Delta)} V_{i,i_{a_{M_{\max}-1}}}| \geq |\sum_{i\in\mathcal{B}_r(i_a;\Delta)} V_{i,i_{a'}}|$ for all $i_{a'} \notin \mathcal{C}_{M,r}(i_a; M_{\max})$.

To ensure that the chosen value of $M_{\max}$ does not begin to artificially modify the dynamics, we can in principle check the convergence of our simulations by varying $M_{\max}$. However, owing to the relatively computationally intensive nature of these calculations, as well as expected variations in the convergence for different parameter regimes, we instead define a proxy quantity to estimate the convergence of our approximation. Specifically, for each cluster we compute

$$R[\mathcal{C}_{M,r}(i_a; M_{\max})] = \frac{\sum_{i_{a'}\in\mathcal{C}_{M,r}(i_a;M_{\max})} \left|\sum_{i\in\mathcal{B}_r(i_a;\Delta)} V_{i,i_{a'}}\right|}{\sum_{i_{a'}\in\mathcal{C}_{M,r}(i_a)} \left|\sum_{i\in\mathcal{B}_r(i_a;\Delta)} V_{i,i_{a'}}\right|}. \tag{S54}$$

The numerator of this expression denotes the sum of the absolute mean couplings of each dipole in the reduced cluster $\mathcal{C}_{M,r}(i_a; M_{\max})$ to the entire buffer zone, whereas the denominator denotes the analogous quantity when summed over each dipole in the full cluster $\mathcal{C}_{M,r}(i_a)$. Thus, $R[\mathcal{C}_{M,r}(i_a; n_{\max})]$ denotes the fraction of the absolute mean couplings that are retained by our reduced cluster, and we take this as figure of merit for determining how small we can afford to make $M_{\max}$ without significantly altering the desired observables.

For our calculations in the main text, we utilize $r$ such that the maximum number of buffer zone sites is $\mathcal{N}_{\max} \leq 20$, and also set a maximum cluster size of $M_{\max} = 15$ (always taking clusters of $M = 8$ for each lattice site in the buffer zone). In Fig. S13, we plot $R[\mathcal{C}_{M,r}(i_a; M_{\max})]$ for both $M_{\max} = 15$, $\mathcal{N}_{\max} = 20$ — as used throughout our simulations in the main text — and $M_{\max} = 12$, $\mathcal{N}_{\max} = 30$. We can see that while the former consistently yields $R \gtrsim 0.90$, the latter leads to much smaller $R \sim 0.5$ for certain tunneling strengths and thus indicates the removal of many dipoles that may contribute significantly to the contrast decay dynamics.

For our final modification, we note that, especially in 2D, it is likely that multiple molecules will initially occupy the buffer zone. For a total number of dipoles $M_{\max}$ in the cluster, of which $N_a$ reside in the corresponding buffer zone, the corresponding dimension of the cluster Hilbert space scales as $\sim \binom{2\mathcal{N}_{\max}}{N_a} \times 2^{M_{\max}-N_a}$; to further limit the size of the corresponding Hilbert space that we must consider, we make the further assumption that *all* dipoles, other than the nucleus of the cluster, remain stationary, so that the corresponding cluster Hilbert space dimension now scales as $\sim 2(\mathcal{N}_{\max} - N_a) \times 2^{M_{\max}-1}$. One potential concern with this approximation is that, particularly in the case of 1D tunneling, these fixed particles will restrict the number of available lattice sites in the buffer zone by blocking access. However, our consideration of extended range tunneling still allows for the nuclear particle to "leapfrog" over any stationary particle, though this will be energetically suppressed by the dipolar interaction.

We note that throughout the main text, we utilize a 3D Gaussian particle distribution matching that described in the experiment, distributed over a lattice of $180 \times 30 \times 150$ sites, with the center of the confining potential located at the center of our lattice along each dimension. Our results for each $E$ field and lattice depth are drawn from $200-2000$ randomly sampled clusters, depending on the associated statistical noise in the mean contrast decay curve. For each lattice depth and $E$ field, we perform EMACE calculations for a range of particle numbers $N_{\text{mol}} \sim 2000 - 4000$.

### C. Comparison to experimental contrast decay traces

To gain further insight into the quantitative disagreement between EMACE simulations and experimental observations of $\kappa$ versus 2D tunneling when $\chi$ = -205 Hz shown in Fig. 4a of the main text, we present contrast decay traces for theory (red lines) and experiment (black circles with fit results as black lines) for similar densities of $n \approx 4\times10^6$ cm$^{-2}$ at four different horizontal lattice depths in Fig. S14. As seen in the panels of Fig. S14, the discrepancy between theory and experiment at intermediate and shallow lattices is a real effect and not simply an artifact from the fitting process to extract $\kappa$. Nonetheless, the EMACE simulations qualitatively capture the peak in $\kappa$ observed experimentally, as shown in Fig. 4a of the main text.

## VII. FURTHER EXPERIMENTAL DATA

### A. Temperature dependence of $\kappa$ in 1D

By compressing the lattice harder in the two tight directions, we increase the temperature in the loose direction. Using a (0,65,65) E$_r$ lattice instead of (0, 30, 30) E$_r$ lattice roughly doubles the temperature. We find that the higher temperature increases the density-dependent contrast decay rates in both the spin-exchange dominated and the Ising dominated cases, consistent with a collisional dephasing picture. We also scanned the dependence of the dephasing on corrugating lattice depth in the (X,65,65) E$_r$ configuration and found that the faster decoherence rates in the shallow lattice cases smoothly connect to the deep lattice case regardless of the tight lattice depth. The results are shown in Fig. S15.

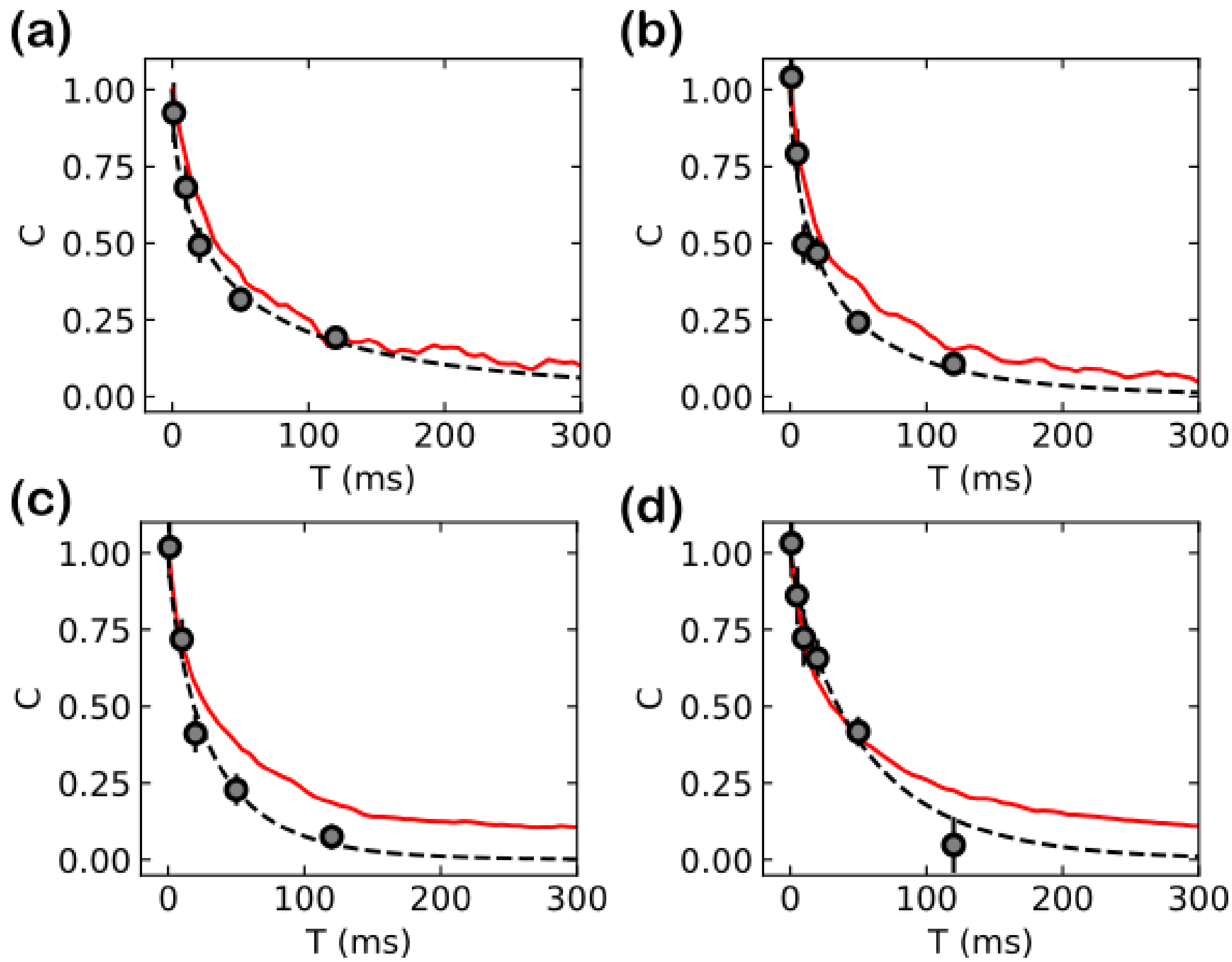

FIG. S14. Contrast time trace comparison between experimental measurements and EMACE simulations for $\chi$ = -205 Hz and $n \approx 4{\times}10^6$ cm$^{-2}$, with black circles experimental data, black dashed line stretched exponential fit to experimental data, and red solid line EMACE simulation for **a)** $V_{L,x} = V_{L,z} = 10$E$_r$ **b)** $V_{L,x} = V_{L,z} = 7$E$_r$ **c)** $V_{L,x} = V_{L,z} = 5$ E$_r$ and **d)** $V_{L,x} = V_{L,z} = 2$ E$_r$. Error bars for experimental points are 1 s.d. from bootstrapping.

### B. $\kappa$ vs 2D tunneling for additional electric fields

We also examined the density-dependent contrast decay for 2D tunneling for two additional electric fields: 2.7 kV/cm corresponding to $\chi$ = -151 Hz and 4.56 kV/cm corresponding to $\chi$ = -77 Hz. The full data set is shown as a 3D plot in $\chi$ and tunneling rate $t$ in Fig. S16. For visual simplicity, we assigned an artificial tunneling rate of 290 Hz for 0 E$_r$ depths. Interestingly, for the $\chi$ = -151 Hz case, in addition to the peak that we observed in the $\chi$ = -205 Hz case, we also observed a dip in contrast when $t$ is around 220 Hz. We note that due to saturation of induced dipole moments, the maximum realizable $\chi$ is approximately the measured $\chi$ = 102 Hz.

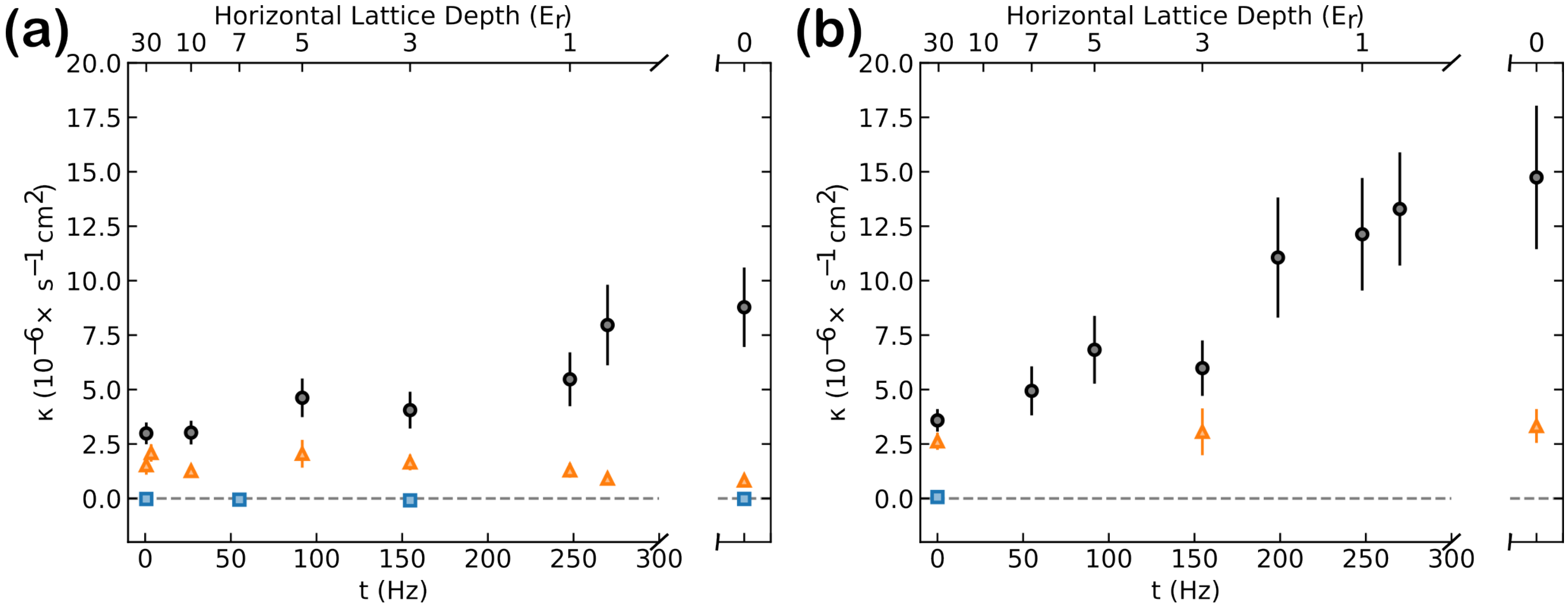

FIG. S15. Extracted density-normalized contrast decay for 1D tunneling at two different temperatures. **a)** Density-normalized contrast decay $\kappa$ versus molecular tunneling rate $t$ in one direction for tight confinements of 30 $E_r$. Black circles, blue squares, and orange triangles present experimental data for $\chi$ = -205, 0, 102 Hz respectively. Error bars are 1 s.e. from linear fits **b)** Same as **a)** but for lattice depths in the tight directions of 65 $E_r$ which roughly doubles the temperature in the loose direction.

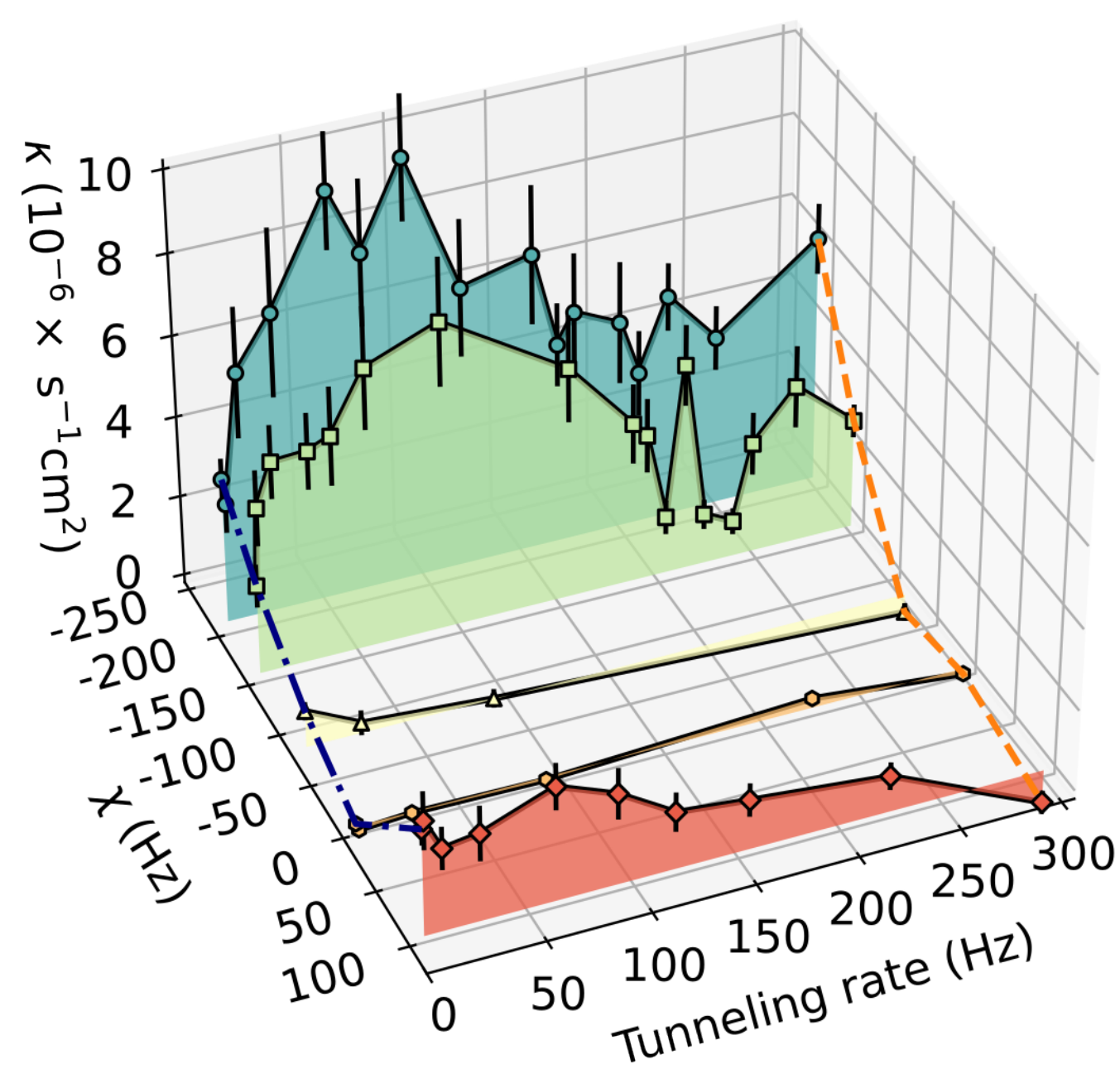

FIG. S16. Extended density-normalized contrast decay for 2D tunneling. Density-normalized contrast decay $\kappa$ versus molecular tunneling rate $t$ for $\chi$ = -205 Hz, -151 Hz, -77 Hz, 0 Hz, and 102 Hz. Points at 290 Hz, traced by orange dashed line, are data taken with horizontal lattices turned off. Error bars are 1 s.e. from linear fits.